\documentclass[12pt]{article}
\usepackage{color}

\usepackage{amsmath,amsfonts,amssymb}
\usepackage{psfrag}
\usepackage{enumerate}
\usepackage{cite}
\usepackage{mathrsfs}
\usepackage{graphicx}
\usepackage{wrapfig}
\usepackage{xcolor}
\usepackage{caption}
\newcommand\blfootnote[1]{%
  \begingroup
  \renewcommand\thefootnote{}\footnote{\hspace{-6mm}#1}%
  \addtocounter{footnote}{-1}%
  \endgroup
}

\newtheorem{theorem}{Theorem}[section]

\newcommand{\cH}{\mathcal{H}}
\newcommand{\la}{\langle}
\newcommand{\ra}{\rangle}
\newcommand{\rb}{\right)}
\newcommand{\lb}{\left(}

\makeatletter\renewcommand\section{\@startsection {section}{1}{\z@}%
                                   {-3.5ex \@plus -1ex \@minus -.2ex}
                                   {2.3ex \@plus.2ex}%
                                   {\normalfont\large\bfseries}}
\renewcommand\subsection{\@startsection{subsection}{2}{\z@}%
                                     {-3.25ex\@plus -1ex \@minus -.2ex}%
                                     {1.5ex \@plus .2ex}%
                                     {\normalfont\bfseries}}

\newcommand{\be}{\begin{equation}}
\newcommand{\ee}{\end{equation}}
\newcommand{\beq}{\begin{eqnarray}}
\newcommand{\eeq}{\end{eqnarray}}

\def\[{\left [}
\def\]{\right ]}

\def\({\left (}
\def\){\right )}

\def\bea{\begin{eqnarray}}
\def\eea{\end{eqnarray}}

\def\r2{\sqrt{2}}

\def\sdiff{S_{\rm diff}}

\newcommand{\nc}{\newcommand}
\nc{\rnc}{\renewcommand}
\nc{\bra}[1]{{\langle#1|}}
\nc{\ket}[1]{{|#1\rangle}}
\nc{\ketbra}[2]{|#1\rangle\!\langle#2|}
\nc{\braket}[2]{\langle#1|#2\rangle}
\nc{\proj}[1]{\left| #1\right\rangle\!\left\langle #1 \right|}
\rnc{\max}{\operatorname{max}}
\nc{\smfrac}[2]{\mbox{$\frac{#1}{#2}$}}
\nc{\tr}{\operatorname{tr}}
\nc{\ox}{\otimes}
\nc{\dg}{\dagger}
\def\ph{\varphi}
\newcommand{\Hmax}{H_{\max}}
\newcommand{\Hmin}{H_{\min}}
\newcommand{\sbr}{\bar{S}}

\newcommand{\lamm}{\lambda_{\max}}

\newcommand{\bbibitem}[1]{\bibitem{#1}\marginpar{#1}}

\newcommand{\figref}[1]{Fig.~\ref{#1}}
\newcommand{\secref}[1]{Sec.~\ref{#1}}

\newcommand{\appref}[1]{Appendix~\ref{#1}}
\newcommand{\nn}{\nonumber}

\def\Label#1{\label{#1}%
  \smash{\hbox to0pt{\raise1ex\hbox{\tiny[#1]}\hss}}}
\def\noLabels{\let\Label=\label}
\def\nobbibitem{\let\bbibitem=\bibitem}

\begin{document}
\noLabels 
\nobbibitem 

\clearpage\thispagestyle{empty}
\begin{center}
{\Large \bf  The Information Theoretic Interpretation \vspace{0.15cm} \\
of the Length of a Curve}

\vspace{7mm}

Bart{\l}omiej Czech$^1$, Patrick Hayden$^1$, Nima Lashkari$^{1,2}$, and Brian Swingle$^1$

\blfootnote{\tt czech, phayden, nimal, bswingle -AT- stanford -DOT- edu}


\medskip\centerline{$^1$ \it Stanford Institute for Theoretical Physics, Stanford University}
\smallskip\centerline{\it 382 Via Pueblo Mall, Stanford, CA 94305-4060, USA}
\bigskip\centerline{$^2$ \it Department of Physics and Astronomy, University of British Columbia}
\smallskip\centerline{\it 6224 Agricultural Road,
Vancouver, B.C., V6T 1W9, Canada}
\end{center}

\vspace{5mm}

\begin{abstract}
\noindent
In the context of holographic duality with AdS$_3$ asymptotics, the Ryu-Takayanagi formula states that the entanglement entropy of a subregion is given by the length of a certain bulk geodesic. The entanglement entropy can be operationalized as the entanglement cost necessary to transmit the state of the subregion from one party to another while preserving all correlations with a reference party. The question then arises as to whether the lengths of other bulk curves can be interpreted as entanglement costs for some other information theoretic tasks. Building on recent results showing that the length of more general bulk curves is computed by the differential entropy, we introduce a new task called constrained state merging, whereby the state of the boundary subregion must be transmitted using operations restricted in location and scale in a way determined by the geometry of the bulk curve. Our main result is that the cost to transmit the state of a subregion under the conditions of constrained state merging is given by the differential entropy and hence the signed length of the corresponding bulk curve. When the cost is negative, constrained state merging distills entanglement rather than consuming it. This demonstration has two parts: first, we exhibit a protocol whose cost is the length of the curve and second, we prove that this protocol is optimal in that it uses the minimum amount of entanglement. In order to complete the proof, we additionally demonstrate that single-shot smooth conditional entropies for intervals in 1+1-dimensional conformal field theories with large central charge are well approximated by their von Neumann counterparts. We also revisit the relationship between the differential entropy and the maximum entropy among locally consistent density operators, demonstrating large quantitative discrepancy between the two quantities in conformal field theories.
We conclude with a brief discussion of extensions and lessons.
\end{abstract}

\setcounter{footnote}{0}
\newpage
\clearpage
\setcounter{page}{1}

\section{Introduction}

Holographic duality (AdS/CFT correspondence) \cite{adscft, witten98} is an equivalence between a $d$-dimensional conformal field theory (CFT) and quantum gravity with asymptotically anti-de Sitter boundary conditions (AdS) in $d+1$ dimensions. Since its discovery in 1997, it has been the focus of a massive body of research\footnote{With over 10,000 citations, Ref.~\cite{adscft} is the most cited paper on the arXiv. See \cite{maldacena2003tasi,hartnoll2009lectures} for pedagogical introductions to the topic, including its applications to condensed matter physics.} driven by diverse theoretical and phenomenological motivations. The AdS/CFT correspondence has been used to model a variety of condensed matter systems, yielding new insights not apparent using the standard techniques of field theory \cite{2005PhRvL..94k1601K,2007PhRvD..75h5020H}. On a more formal level, famous puzzles of quantum gravity, including its unitarity and non-perturbative definition, have been addressed and arguably solved by adverting to the field theory side of the duality (see, e.g. \cite{myreview, tedsessay}).

This paper focuses on another foundational application of the AdS/CFT correspondence -- the goal of understanding the fundamental constituents of space-time. Note that the duality relates a lower-dimensional field theory to a higher-dimensional theory of quantum gravity. The extra dimension is said to be emergent: on the field theory side it is not directly visible, but becomes apparent only when we discuss an appropriate set of quantities. In this way, the holographic duality is a toy model for how a geometric spacetime may arise from an amorphous collection of quantum gravity degrees of freedom, which lack an {\it a priori} spatial organization. In order to reap this benefit of holography, we must understand quantitatively how the lower-dimensional field theory gives rise to the extra dimension present in the gravitational space-time. The last years have made it increasingly clear that the right language for this problem involves quantum information theory. The conceptual link between information theory and the geometry of a holographic spacetime is the subject of the present paper.

Until recently, the understanding of the extra dimension (usually called radial) in holography had been mostly qualitative. It was understood early on that small distance physics in the field theory controls large radial scales on the gravity side, a rule of thumb known as the UV-IR connection \cite{uvir}. Consequently, the radial scale was conjectured to be related to a renormalization group (RG) scale in field theory \cite{Akhmedov:1998vf, Balasubramanian:1999jd, deBoer:1999xf}. But the RG scheme implementing the radial evolution in gravity has never been explicitly identified (see \cite{Balasubramanian:2012hb, Jackson:2013eqa} for recent progress). The first truly quantitative advance -- one whose consequences continue to be explored -- came in 2006. The Ryu-Takayanagi proposal \cite{rt1, rt2} posits that areas of minimal surfaces on a static slice of anti-de Sitter space compute entanglement entropies of spatial regions in field theory. To appreciate the significance of this proposal, recall that a combination of entanglement entropies called mutual information bounds the connected correlator of any two observables applied in two spacelike separated regions \cite{entcorr}. This means that entanglement entropies organize the correlations in a quantum state as a function of distance or scale. In effect, the Ryu-Takayanagi proposal posits that the amount of correlation up to a given scale $\mu_0$ in field theory can be represented in anti-de Sitter space as a minimal surface, which spans different radial slices down to some minimal scale $R_0$ that depends on $\mu_0$. Amazingly, this geometric representation is quantitatively accurate.

The Ryu-Takayanagi proposal underscores the centrality of information theory to the emergence of a holographic spacetime. For example, it clarifies why spacetimes with horizons correspond to mixed states of the field theory and why a black hole with two asymptotic regions maps to the thermofield double state \cite{Horowitz:1998xk, Balasubramanian:1998de, Maldacena:2001kr}. Moreover, in a thought experiment in which we disentangle two regions of field theory by hand, the holographic spacetime pinches off into disconnected components \cite{markessay, rqg}. (When more than two regions are considered, however, mutual information can be zero between two connected regions of spacetime~\cite{balasubramanian2014multiboundary}.) More quantitatively, representing entanglement entropies with minimal surfaces is automatically consistent with the strong subadditivity of entropy as a consequence of  the geometric properties of anti-de Sitter space \cite{ssaproof}. Indeed, the strong subadditivity inequality plays a fundamentally geometric role in the holographic construction: in the AdS$_3$/CFT$_2$ context, for instance, it underlies the triangle inequality in AdS$_3$ \cite{lampros} and reduces the $c$-theorem in CFT$_2$ to Lorentz invariance \cite{ssalorentz}. More intricate relations among minimal surfaces have been used to identify special properties of states in holographic field theories, including the monogamy of mutual information~\cite{monogamy}.
This web of connections has motivated several authors to conjecture that a spacetime should be identified with (or defined as) a geometric encoding of field theory correlations organized by scale \cite{briansessay, marks1st, markessay, brians2nd, bianchimyers, tomjuan, myerssmolkin, holeentropy, erepr, xiaoliang, complexity}. If so, every geometric construct in a holographic spacetime should have a meaning in information theory. The present paper interprets in information theoretic terms one of the most basic geometric objects: the length of a convex curve.

We work primarily in pure, three-dimensional anti-de Sitter space, which is dual to the vacuum of a two-dimensional conformal field theory. Some of our results are more general, but we defer a discussion of the generality to Sec.~\ref{disc}. A key technical fact borrowed from \cite{holeography} -- and a starting point of our work -- is that the length of a convex curve in AdS$_3$ can be written as a linear combination of lengths of minimal curves, that is geodesics. By virtue of the Ryu-Takayanagi proposal, the latter compute entanglement entropies of intervals in the dual CFT$_2$. In consequence, the length of a convex curve can expressed as:
\begin{align}
\frac{\rm length}{4G} & = \int \Big(S\big(I(x)\big) - S\big(I(x) \cap I(x-dx)\big) \Big)
\nonumber \\ & =
\int S\big( I(x) - I(x-dx)\, |\,  I(x) \cap I(x-dx) \big) \equiv \sdiff
\label{mainintro}
\end{align}
Here $S(\cdot)$ denotes the entanglement entropy and $I(x)$ is a one-parameter family of boundary intervals, each centered at $x$, which determine the shape of the curve in question. The integral expressions in (\ref{mainintro}) were called {\bf differential entropy} in \cite{holeography}. Note that the second line of (\ref{mainintro}) involves only conditional entropies in field theory. This suggests that the length of a curve may be interpreted in information theory as the entanglement cost of a merging task \cite{naturepaper,horodecki2007quantum}. The details of such a task are the subject of Sec.~\ref{diff-entropy}.

Suppose that Alice, who controls a CFT from the outside, wishes to send the state on an interval $I$ to Bob. By sending we mean transferring the entanglement between $I$ and $I^c$, the complement of $I$ in the CFT, to another system controlled by Bob. Because the key object being transferred is the entanglement, classical communication is considered free. The cost refers to an inherently quantum resource -- the entanglement between Alice and Bob's systems, which is used in the process of sending the state. A natural choice of currency for quantifying the cost is Bell pairs, which are initially shared by Alice and Bob.

One way for Alice to send the state to Bob is to compress the state on $I$ to $S(I)$ binary degrees of freedom \cite{schumacher} and then to teleport them \cite{teleport} to Bob. This will use up exactly $S(I)$ units of the entanglement currency. As such, sending the state gives an operational meaning of the entanglement entropy and, by virtue of the Ryu-Takayanagi proposal, of the length of a geodesic in AdS$_3$. To make contact with (\ref{mainintro}), we now imagine that the merging is done in steps indexed by $x$, such that both Alice and Bob may only act on interval $I(x)$ at step $x$. The details of this {\bf constrained merging protocol} are given in Sec.~\ref{diff-entropy}.\footnote{The constrained merging protocol is relevant to closed curves with two endpoints on the asymptotic boundary. To interpret the length of a closed convex curve, in Sec.~\ref{swapprotocol} we introduce its close cousin, the {\bf constrained swapping protocol}.} A key point is that the optimal cost of sending a state subject to the locality restrictions imposed by the intervals $I(x)$ is exactly the length of the curve given in \eqref{mainintro}. The proof of the optimality of (\ref{mainintro}) is given in Sec.~\ref{optimality}. Sec.~\ref{sec:single-shot} establishes an important technical point crucial to identifying the constrained merging cost with the differential entropy, specifically that smooth min- and max- entropies are well-approximated in CFT's with large central charge by the von Neumann entropy.

From a geometric viewpoint, the intervals $I(x)$ determine the shape of the bulk curve. In a traditional view of the radial direction as an RG scale in field theory, we could think of $I(x)$ as determining a spatially dependent cutoff in the CFT. The present paper offers an alternative view, which may serve as a gateway toward a quantitative formulation of holographic RG. We think of $I(x)$ as restricting the class of operators, which are available to external agents manipulating the state. This can be viewed as a spatially dependent restriction of the class of operators of the field theory, which excludes IR-sensitive observables. The restrictions may only increase the cost of sending a state; when we lift the restrictions, the cost becomes the entanglement entropy. This gives an information theoretic interpretation of the definition of a geodesic as the shortest curve connecting two points.

\section{Differential entropy and constrained state merging}
\label{diff-entropy}
In this section we focus on pure three-dimensional anti-de Sitter space (AdS$_3$). Our results apply in other asymptotically AdS$_3$ geometries and in higher-dimensional holographic spacetimes, but they are subject to a number of technical caveats. We discuss the generality of our results in Sec.~\ref{disc}.

We start with the metric on the Poincar\'e patch of AdS$_3$:
\begin{equation}
ds^2 = - \frac{R^2}{L^2}\, dT^2 + \frac{L^2}{R^2}\, dR^2 + R^2 d\tilde{x}^2.
\label{ads3poincare}
\end{equation}
We assume that this geometry arises as the dual description of the vacuum state of a conformal field theory (CFT) living on its asymptotic boundary -- that is on an infinite line cross time. We denote the transversal coordinate in the bulk as $\tilde{x}$, in contrast to $x$, which we reserve for the spatial coordinate on the boundary.

The Ryu-Takayanagi proposal \cite{rt1, rt2} relates the entanglement entropy of an interval $I = (-a/2, a/2)$ in the CFT to the length of the spacelike geodesic, which asymptotes to the endpoints of $I$:
\begin{equation}
S(I) \equiv S(a) = \frac{c}{3} \log \frac{a}{\mu} = \frac{\textrm{length of geodesic connecting $\tilde{x}=\pm a/2$ at $R=L^2/\mu$}}{4G}
\label{vacent}
\end{equation}
The quantity $\mu$, which is a UV cutoff in the CFT, also defines an IR cutoff $L^2/\mu$ on the dual gravity side, which regulates the otherwise infinite length of the geodesic. Eq.~\eqref{vacent} relies on the Brown-Henneaux relation $c = 3L/2G$, which fixes the central charge of the 1+1-dimensional CFT in terms of the curvature scale $L$ of the dual AdS$_3$ in Planck units ($G$ is Newton's constant) \cite{brownhen}.

Ref.~\cite{holeography} (see also \cite{roblast, xi, entwinement, Wienthesis, robproof, lampros}) showed how to use relation (\ref{vacent}) to give a boundary computation of the length of an arbitrary differentiable curve on a constant time slice in geometry (\ref{ads3poincare}). Given a convex\footnote{With respect to metric~(\ref{ads3poincare}). A curve is convex if none of the geodesics tangent to it intersect it again.} curve $R = R(\tilde{x})$, for every point $\tilde{x}$ one finds the geodesic that is tangent to the curve at $\tilde{x}$. The endpoints of the geodesic lie on the asymptotic boundary, so they select a boundary interval. We shall refer to this interval as $I(x)$, where $x$ is the midpoint of the interval.
Likewise, we denote the linear size of $I(x)$ by $a_I(x)$.
Note that $x$ depends on $\tilde{x}$ (the tangency point in the bulk) but is not equal to it. The construction is illustrated in the case of a geodesic curve in \figref{picgeodesic}a and a nongeodesic curve in \figref{picnongeodesic}a.

The length of the curve is then given by the formula:\footnote{For general open curves in the bulk of (\ref{ads3poincare}), the integral must be supplanted with boundary terms. We concentrate on curves with endpoints at infinity, to which they are not relevant.}
\begin{align}
\frac{\rm length}{4G} & = \int \Big(S\big(I(x)\big) - S\big(I(x) \cap I(x-dx)\big) \Big)
\nonumber \\ & =
\int S\big( I(x) - I(x-dx)\, |\,  I(x) \cap I(x-dx) \big) \equiv \sdiff
\label{main}
\end{align}
Note that the integrand in \eqref{main}, or rather the first nonvanishing term in its Taylor expansion, is a one-form, so the integral is well-defined. The right hand side was called ``differential entropy'' in \cite{holeography}, because the integrand can be expressed in terms of $dS/da$. For the purposes of this paper, however, it is most practical to work directly with expression (\ref{main}), which involves conditional entropies. By definition, the conditional entropy $S(A|B)$ of two disjoint subsystems $A$ and $B$ is the difference $S(AB)-S(B)$.

To state our result, it is useful to introduce a little extra notation and discretize \eqref{main}. Let $x_j = -a/2 + j \cdot a/N$ and define $A_j = I(x_j ) - I(x_{j-1})$ (hinting at Alice) and $B_j = I(x_j) \cap I(x_{j-1})$ (hinting at Bob). Then \eqref{main} becomes:
\begin{equation}
\frac{\text{length}}{4G}= \lim_{N \rightarrow \infty} \sum_{j=1}^N
	\left[ S \left(I(x_j) \right) - S\left( I(x_j) \cap I(x_{j-1} ) \right) \right]
= \lim_{N \rightarrow \infty} \sum_{j=1}^N
	S( A_j | B_j ). \label{eqn:discretized}
\end{equation}
In order to interpret \eqref{eqn:discretized} suppose that two agents, Alice and Bob, each hold a system described by a CFT, which they can manipulate from outside. For example, the system may be a one-dimensional spin lattice at a quantum phase transition that is sitting in Alice's laboratory; Bob's lab contains an isomorphic lattice. Initially Alice and Bob's systems are not entangled, i.e. their states factorize. Alice's goal is to ``teleport''~\cite{teleport} the state of interval
\begin{equation}
I = \cup_{j=1}^N I_j = \cup_{j=1}^N A_j
\end{equation}
of her CFT to Bob.\footnote{In the last line, we set $A_1 = I(x_1)$ as a special case.} That is, using only Alice-Bob Bell pairs $\smfrac{1}{\sqrt{2}}( \ket{00} + \ket{11} )$ and classical communication plus local operations (we discuss the locality constraints below), Alice and Bob will prepare a state in Bob's lab equal to Alice's original state on $I$ (up to isomorphism) \emph{and} purifying $I^c$, the complement of $I$ in Alice's system.

A crucial role in interpreting lengths of curves is played by the locality restrictions, which constrain the type of operations Alice and Bob can perform. If we allow Alice and Bob to each act on the whole of their respective intervals $I$, their task reduces to standard teleportation \cite{teleport}, whose cost is famously given by the entropy of the state to be teleported, $S(I)$. We wish to consider a situation, in which Alice and Bob are subject to tighter locality constraints. The procedure will act in $N$ discrete steps and at the $j^{\rm th}$ step Alice and Bob are allowed to act \emph{only} on their respective intervals $I(x_j)$. ($N$ will ultimately be allowed to go to infinity as the UV cutoff $\mu$ goes to zero and the central charge to infinity.)
We will consider all possible procedures that Alice and Bob could use to ``merge'' Alice's $I$ to Bob, subject only to the prescribed constraints. (A mathematically precise definition of constrained merging can be found in \secref{optimality}.) Among all such procedures, the minimal number of Bell pairs  is asymptotically given by \eqref{eqn:discretized}, the length of the curve in Planck units.

\subsection{A geodesic: the cost of sending a state}
\label{explgeodesic}

\begin{figure}[t!]
\centering
\raisebox{4cm}{a)}\includegraphics[width=.5\textwidth]{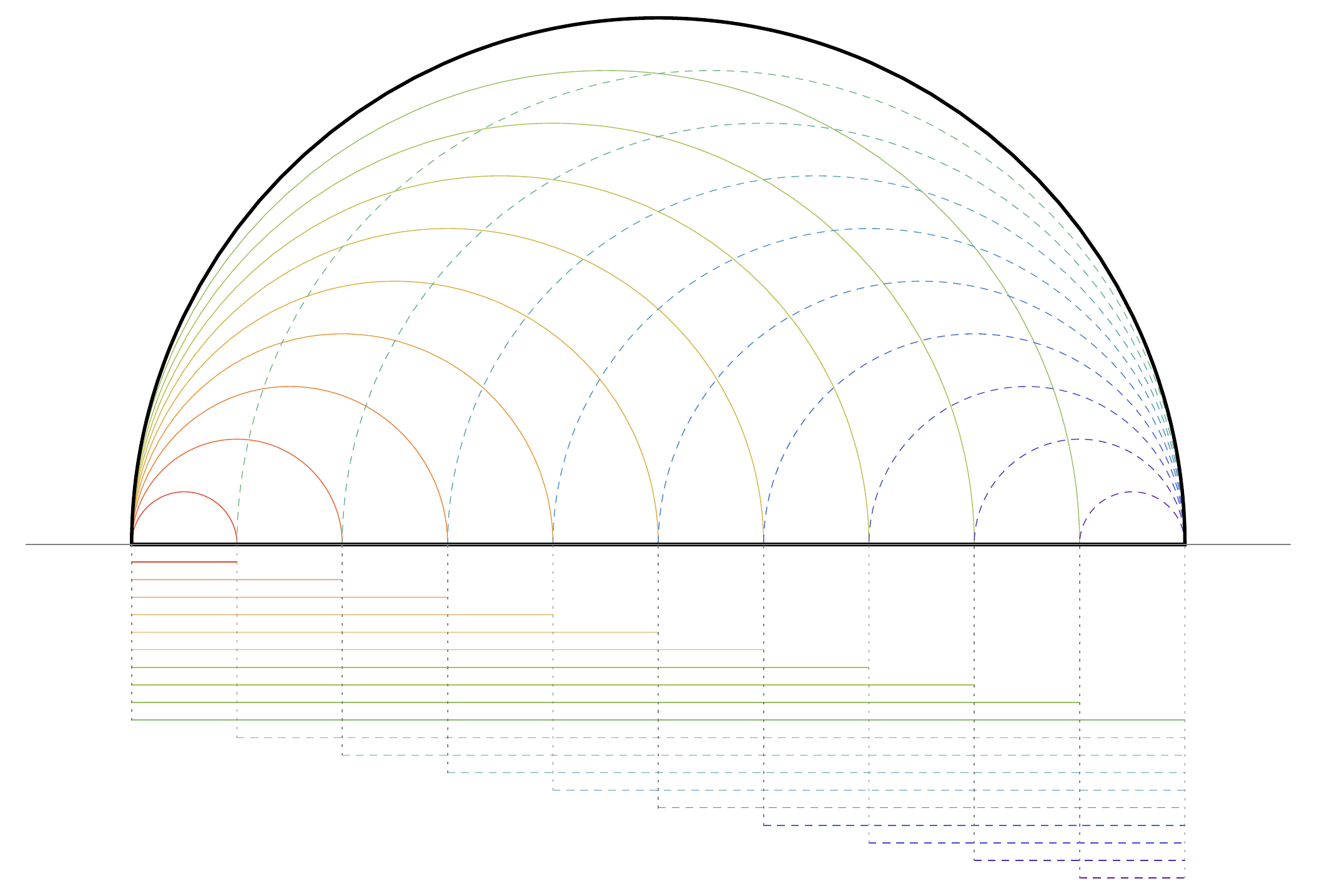}
\raisebox{4cm}{b)}\includegraphics[width=.35\textwidth]{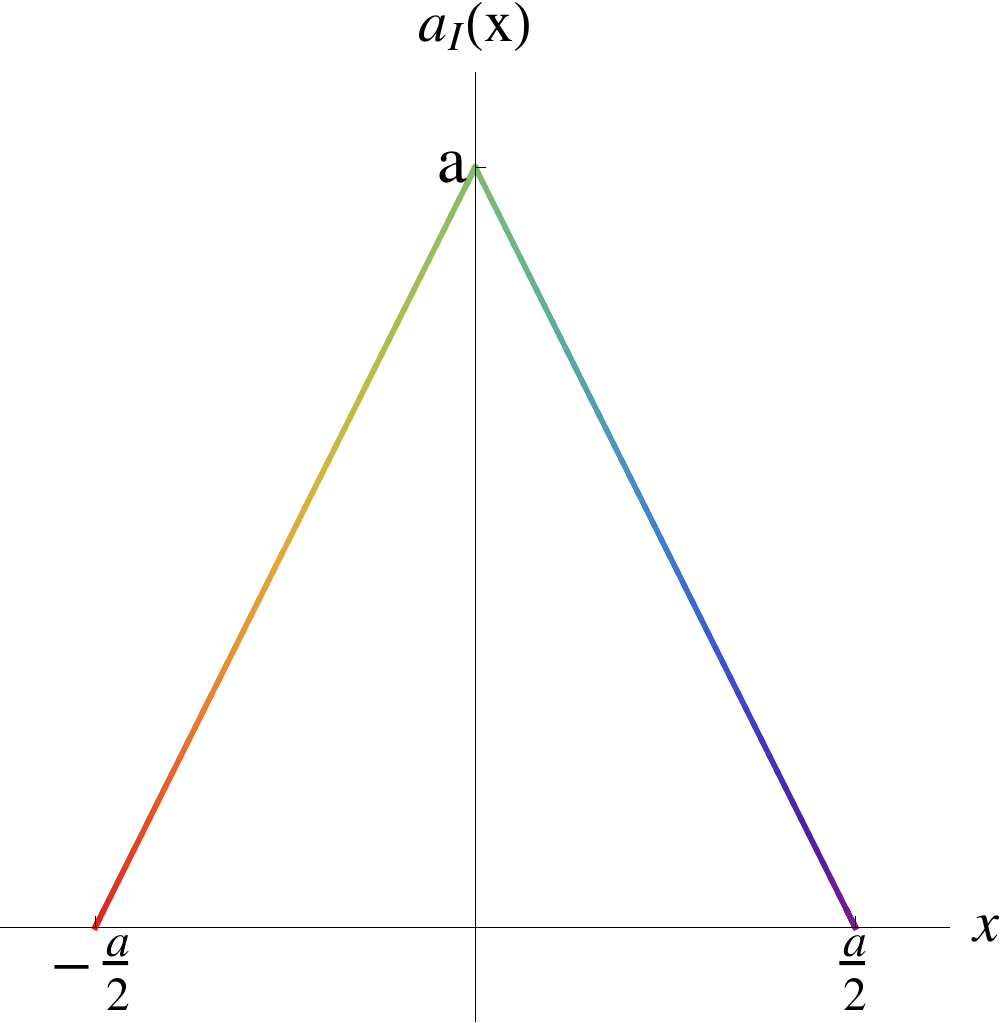}
\caption{a) Geodesic $g_I$, which subtends a boundary interval $I$ (black) and geodesics, which are tangent to $g_I$ on the boundary along with the corresponding boundary intervals $I(x)$ (color). The dashed geodesics contribute zero to integral (\ref{main}). b) $a_I(x)$, the linear size of the interval $I(x)$ centered at $x$.}
\label{picgeodesic}
\end{figure}

We begin with the interpretation of the length of a spacelike geodesic $g_I$. For definiteness, suppose the geodesic subtends the interval $I = (-a/2, a/2)$ on the boundary. According to \eqref{vacent}, the (IR-regulated) length of $g_I$ equals $S(I)$, the (UV-regulated) entanglement entropy of the interval $I$.

As a first step, we must find the set of intervals $I(x)$ such that geodesics subtending $I(x)$ are tangent to $g_I$. The task seems trivial, because at every point on $g_I$ the tangent geodesic is $g_I$ itself. But this conclusion holds only in the bulk; on the asymptotic boundary distinct geodesics become tangent to one another if their asymptotic endpoints coincide; see Fig.~\ref{picgeodesic}. Thus, the sequence of intervals $I(x) = (-a/2, a/2 + 2x)$ (for $-a/2 \leq x \leq 0$) and $I(x) = (2x - a/2, a/2)$ (for $0 \leq x \leq a/2$) satisfies the tangency condition. (While we will use the notation of infinitesimals for simplicity, the reader should remember that we are always describing discretized expressions and processes, both because the CFT has a UV cut-off $\mu$ and because the procedure we implement will take place in finite steps.)

Let us consider the entanglement cost of merging $I$ to Bob subject to the constraints described above. For early values of $j$, corresponding to red and orange intervals in Fig.~\ref{picgeodesic}, Alice and Bob are only permitted to act on the left side of $I$, while for blue and purple intervals they can only act on the right. For $x_j =0$, however, corresponding to the full length green interval, they have access to all of $I$. So Alice could simply compress the state of $I$~\cite{schumacher} in the $x_j=0$ step and teleport it to Bob, who would then decompress on his end. At all other steps, Alice and Bob would do nothing. The entanglement cost, postponing until later issues of single-shot versus von Neumann entropies and approximation, would be $S(I)$.  This is, of course, the familiar interpretation of the entropy as the effective number of Bell pairs required to faithfully compress and teleport the state of $I$ without any constraints at all.

\subsection{A non-geodesic curve: the cost of sending a state with constrained merging}
\label{explnongeodesic}

\begin{figure}[t!]
\centering
\raisebox{6cm}{a)}\includegraphics[width=.5\textwidth]{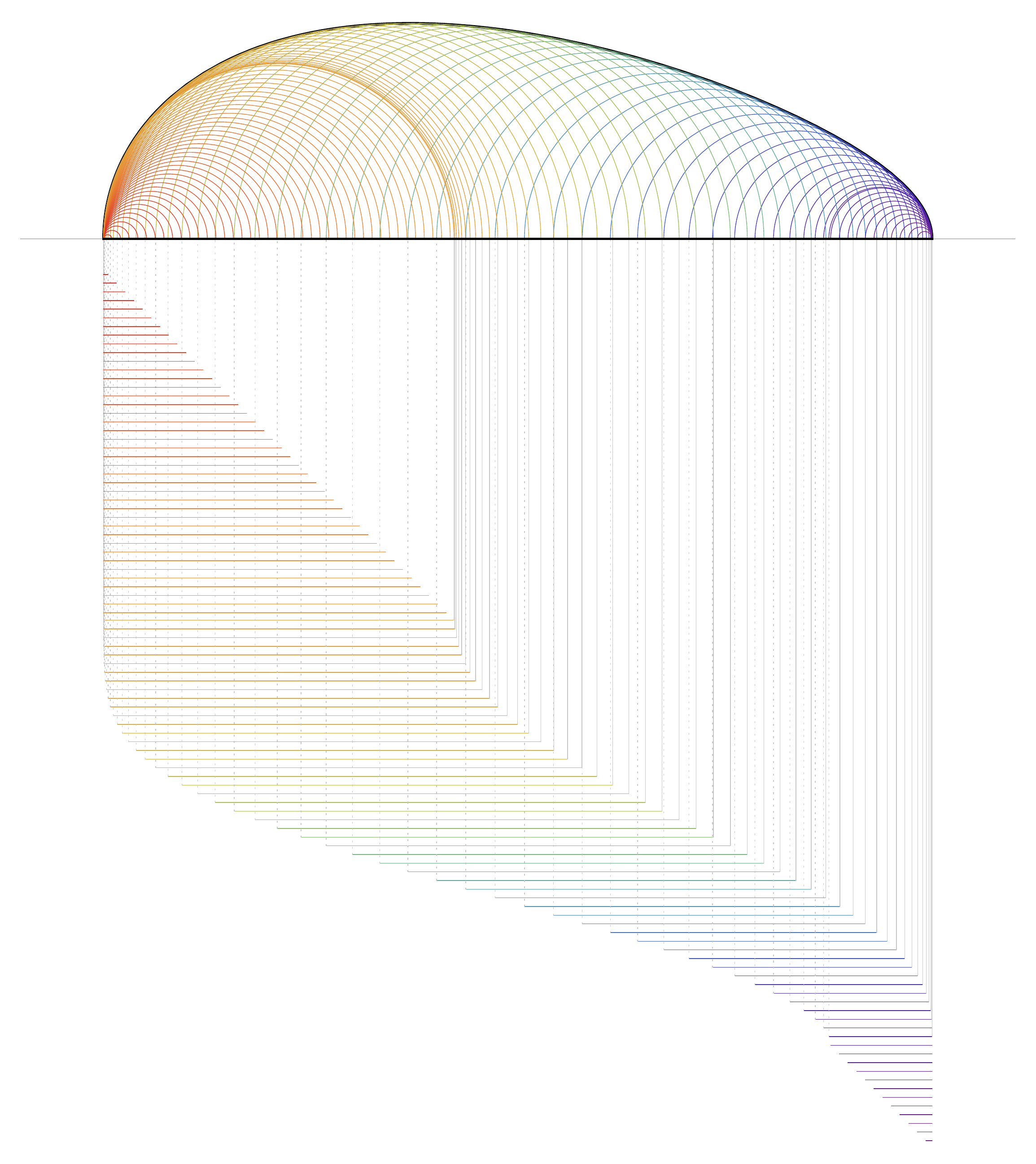}
\raisebox{6cm}{b)}\raisebox{2cm}{\includegraphics[width=.35\textwidth]{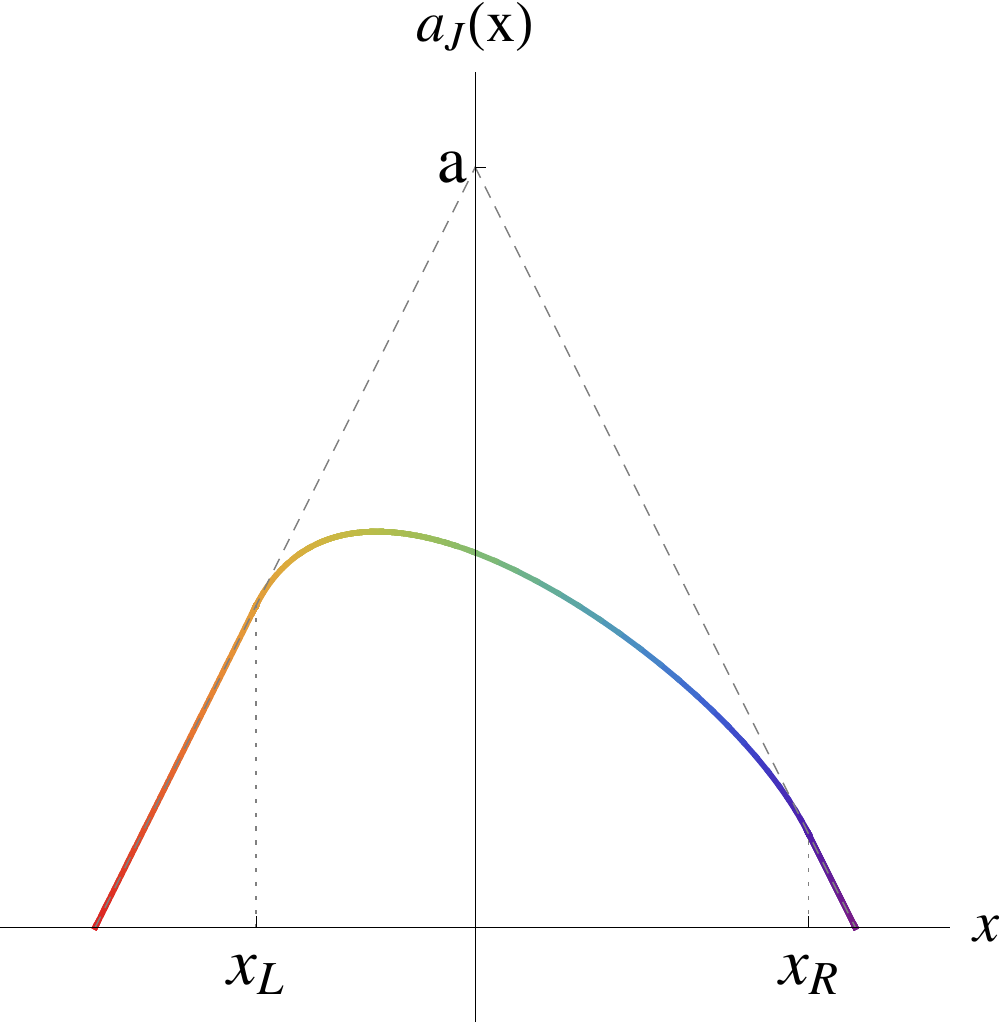}}
\caption{a) A curve, which asymptotes to the geodesic $g_I$. We have marked in color the geodesics tangent to the curve and the boundary intervals $J(x)$, which they subtend. b) a plot of $a_J(x)$ -- the length of the intervals $J(x)$ as a function of the centerpoint $x$. If the curve asymptotes to the geodesic $g_I$ then $a_J(x)$ must agree with $a_I(x)$ (shown for comparison in dashed gray) outside some interval $(x_L, x_R)$ \cite{lampros}.}
\label{picnongeodesic}
\end{figure}

Now consider a smooth, convex curve $R = R(\tilde{x})$, which asymptotes to the endpoints $\pm a/2$ of the interval $I$; see Fig.~\ref{picnongeodesic}. We again start by finding the geodesics, which are tangent to the curve at every $-a/2 \leq \tilde{x} \leq a/2$. These geodesics select a sequence of boundary intervals, which we denote $J(x)$ to distinguish them from the intervals discussed in the previous subsection. It will be useful to introduce a special notation for the size of the interval $J(x)$: we call this $a_J(x)$. Note that for the geodesic we have $a_I(x) = a - 2|x|$ for the function giving the lengths of the intervals $I(x)$, as illustrated in \figref{picgeodesic}. For our nongeodesic curve $R = R(\tilde{x})$ that asymptotes to $g_I$, $a_J(x)$ has the following properties \cite{lampros}:
\begin{enumerate}
\item It equals $a_I(x)$ outside a certain interval $(x_L, x_R) \subset I$.
\item It is everywhere differentiable, with $-2 \leq da_J/dx \leq 2$.
\end{enumerate}
An example of such an $a_J(x)$ is shown in Fig.~\ref{picnongeodesic}b. According to \eqref{main}, the length of the curve is given by:
\begin{align}
\frac{\rm length}{4G} &= %
\int_{x=-a/2}^{x=a/2} S\big( J(x) - J(x-dx)\, |\,  J(x) \cap J(x-dx) \big)
= \lim_{N\rightarrow\infty} \sum_{j=1}^N S(A_j | B_j),
\label{nongeodlength}
\end{align}
where now $A_j = J(x_j) - J(x_{j-1})$ and $B_j = J(x_j) \cap J(x_{j-1})$.

In contrast to the geodesic case, for the nongeodesic curve of \figref{picnongeodesic}a, none of the intervals $J(x_j)$ spans all of $I$. The simple-minded strategy of Sec. \ref{explgeodesic} therefore cannot succeed. Instead, Alice and Bob will act non-trivially in each interval $J(x_j)$. Specifically, in the $j$th step, Alice will merge $A_j$ to Bob. Since $A_j \subseteq J(x_j)$, Alice's actions are consistent with the constraint. Moreover, by the $j$th step, Bob will already have reconstructed the entire interval $\cup_{i=1}^{j-1} J(x_i)$, of which the rules give him access only to the portion intersecting $J(x_j)$, namely $J(x_{j-1}) \cap J(x_j) = B_j$.

The question becomes then, what is the cost of each incremental step? A celebrated result in quantum information theory is that the number of Bell pairs required to merge $A_j$ provided Bob has access to $B_j$ is $S(A_j | B_j)$, again ignoring approximations for the time being~\cite{naturepaper,horodecki2007quantum}. By \eqref{nongeodlength}, the length of the bulk curve will therefore be approximated by $4 G$ times the number of Bell pairs required to merge $I$ to Bob, subject to the locality constraints.

For readers unfamiliar with state merging, it can be helpful to keep some simple examples in mind to motivate the appearance of the conditional entropy. If the state is a Bell pair shared between $A$ and a third system $R$, then merging $A$ to Bob is just teleportation of $A$ and the cost in Bell pairs is indeed $S(A|B) = S(A) = 1$.

If the initial state is instead a GHZ state $\ket{\psi}_{ABR} = \smfrac{1}{\sqrt{2}} ( \ket{000}_{ABR} + \ket{111}_{ABR})$, however, then Bob has a head start in the form of some correlation with Alice so the cost should be reduced. In fact, since $S(A|B) = S(AB) - S(B) = 1 -1 = 0$, merging should be possible without any entanglement at all.
Let's see how this is done. Alice could measure $A$ in the basis $\ket{\pm} = \smfrac{1}{\sqrt{2}} ( \ket{0} \pm \ket{1} )$. Conditioned on detecting outcomes $\ket{+}$ and $\ket{-}$, the state on $BR$ will be $\smfrac{1}{\sqrt{2}} ( \ket{00}_{BR} + \ket{11}_{BR} )$ or $\smfrac{1}{\sqrt{2}} ( \ket{00}_{BR} - \ket{11}_{BR} )$, respectively. Both of these states are maximally entangled with $R$ and can, therefore, be transformed into $\smfrac{1}{\sqrt{2}} ( \ket{000}_{A'B'R} + \ket{111}_{A'B'R} )$ by a local operation in Bob's laboratory. Explicitly, in the first case, Bob can apply the isometry $\ket{j}_B \mapsto \ket{jj}_{A'B'}$ and in the second case, he can apply $\ket{j}_B \mapsto (-1)^j \ket{jj}_{A'B'}$. Since no Alice-Bob entanglement is consumed or produced by the protocol, the cost is precisely zero.

The method used in the general case is in spirit of the GHZ example above and involves performing a random incomplete measurement on $A$ just fine-grained enough to approximately destroy all correlation between $R$ and $A$. The reader can consult \cite{horodecki2007quantum} for a detailed description.

\subsection{Geodesics revisited: merging scale-by-scale}
\label{scalebyscale}

Let us return to the geodesic case in order to study in more detail how the entanglement entropy is recovered from the differential entropy formula.

If we substitute the sequence of intervals from \figref{picgeodesic}b into \eqref{main}, we obtain:
\begin{equation}
\sdiff = \int_{x = -a/2}^{x=0} \Big( S\big(I(x)\big) - S\big(I(x-dx)\big)\Big)  = S\big(I(0)\big) - S\big(I(-a/2)\big) = S(I)
\label{geodlength}
\end{equation}
This is because for  $-a/2 \leq x \leq 0$ we have $I(x-dx) \subset I(x)$; for $0 \leq x \leq a/2$ the integrand in \eqref{main} vanishes because there $I(x) \subset I(x-dx)$, so $A_j = \emptyset$. The term $I(-a/2)$ acts as a UV regulator; it reproduces \eqref{main} exactly if we cut off the integral at $x = -a/2 + \mu/2$.

In the notation that $S(I) = S(a)$, \eqref{geodlength} can be written as:
\begin{equation}
\sdiff = \int_{x = -a/2}^{x=0} \big(S(a+2x) - S(a+2x -2dx) \big)= \int_{a_I = 0}^{a_I = a} S(da_I | a_I - da_I) = S(a)
\label{geodbyscale}
\end{equation}
Changing the variable of integration to $a_I = a+2x$ highlights the way in which the differential entropy formula recovers ordinary entanglement entropy. It assembles it from successive pieces, which incorporate the entanglement at incrementally larger scales.

According to the rules of constrained merging, in the geodesic case it is permissible for Alice to send the interval $I$ to Bob all at once, as described in Sec.~\ref{explgeodesic}. She is not required to, however. Instead, she could use the incremental procedure described for general curves in Sec.~\ref{explnongeodesic} and the cost would be the same.
In the incremental procedure, she starts out by sending Bob the most ultraviolet data on $I$ -- the state of the smallest sensible interval tucked at the left endpoint of $I$, where ``smallest'' means roughly comparable to the UV cutoff in the CFT, in a sense that will be made precise in \secref{sec:single-shot}. In the next steps, she will send data necessary to recover the state on successively larger intervals. If Bob has the state on $I(x-dx)$, the number of ebits necessary for him to recover the state on $I(x)$ is $S\big(I(x)\big) - S\big(I(x-dx)\big) = S\big(I(x) - I(x-dx) \, | \, I(x-dx)\big)$, which is the integrand in \eqref{geodlength}.  The vanishing of the integrand in \eqref{main} for $0 \leq x \leq a/2$ expresses the fact that once Bob has the state on $I(x=0) = I$, there is nothing more to learn and thenceforth the cost is zero.

\subsection{Minimality of geodesics: the most efficient merging protocol}
\label{min-geodesic}

This perspective is also helpful for interpreting the length of the nongeodesic curve of \figref{picnongeodesic}. Because $a_J(x) = a_I(x)$ for $x < x_L$, as shown in the righthand side of the figure, the incremental protocol for the nongeodesic curve will begin in exactly the same way as for the geodesic, sending the most UV information near the point $-a/2$ first and then the data required to reconstruct the state of successively longer intervals $(-a/2,x)$. Once the interval reaches $I(x_L)$, however, the protocols diverge. In the nongeodesic case, Alice and Bob are constrained to act over shorter distances than in the geodesic case so they cannot access the IR information as efficiently. The result is an increased entanglement cost in the merging protocol, which matches the difference in length between the two curves.

This gives an information theoretic interpretation of the definition of the geodesic as the shortest path between two points. Any other curve with the same endpoints as $g_I$ will select a different $a_J(x)$, which corresponds to a \emph{constrained} merging protocol. Any restriction imposed on the merging protocol can only increase the cost of communication. Consequently, any other path connecting the same endpoints on the boundary is longer than a geodesic.

To see this algebraically, we first need to obtain an analogue of \eqref{geodbyscale} for the nongeodesic case:
\begin{align}
\sdiff = &
\int_{x = -a/2}^{x=a/2} \Bigg(S\big(a_J(x)\big) - S\left(\frac{a_J(x) + a_J(x-dx)}{2} - dx \right) \Bigg)
\nonumber \\ = &
\int_{x = -a/2}^{x=a/2} S\left(\frac{a_J(x) - a_J(x-dx)}{2} + dx \,\Big|\, \frac{a_J(x) + a_J(x-dx)}{2} - dx \right)
\nonumber \\ = &
\int_{x = -a/2}^{x=a/2} S\left(\frac{da_J+2dx}{2} \, \Big| \, a_J(x) - \frac{da_J + 2dx}{2}\right)
\label{nongeodbyscale}
\end{align}

\begin{figure}[t!]
\centering
\includegraphics[width=.5\textwidth]{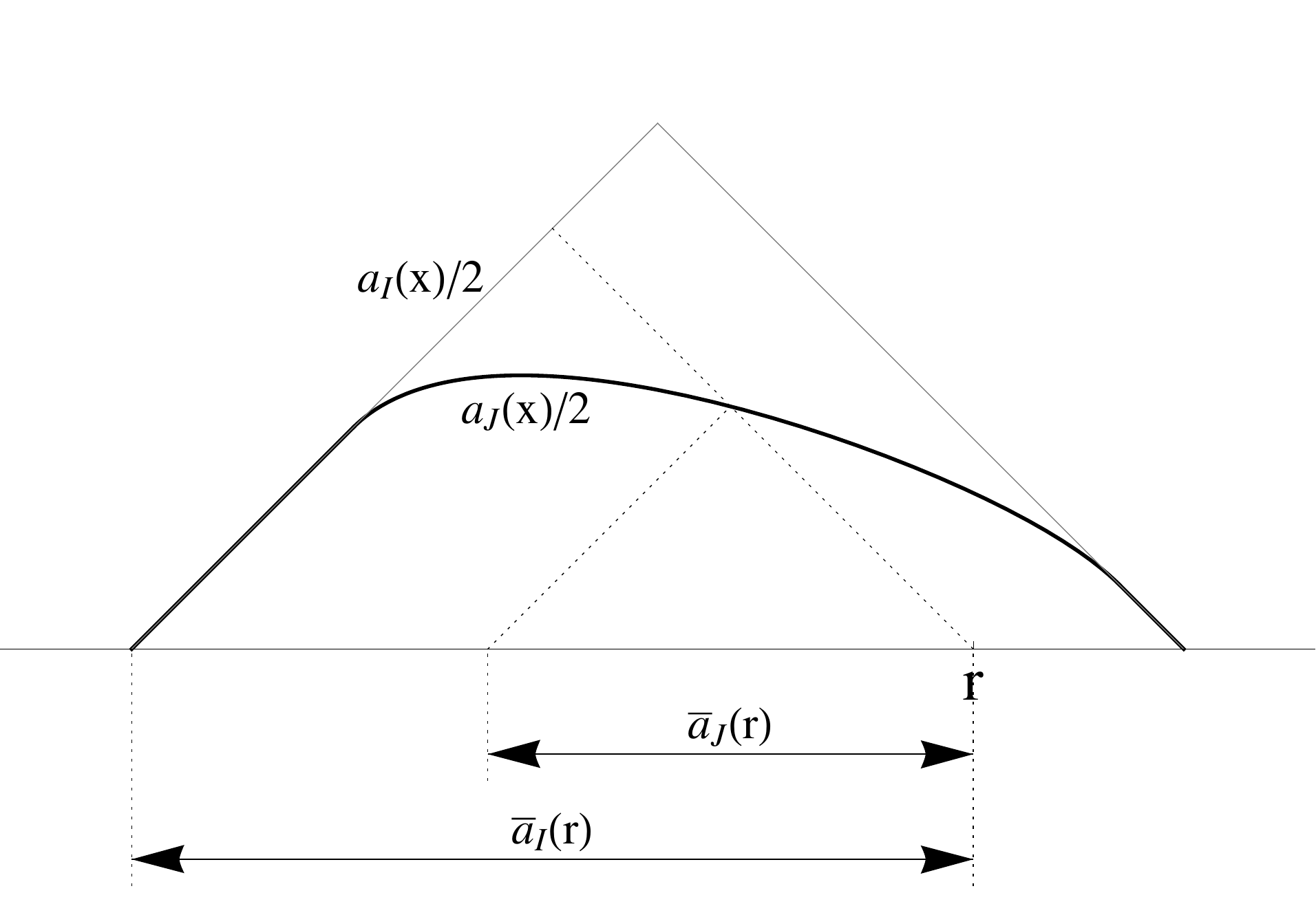}
\caption{The way to read off $\bar{a}_{I,J}(r)$ from the plot of $a_{I,J}(x)/2$.}
\label{abars}
\end{figure}

With the formula in hand, subtract \eqref{geodbyscale} from \eqref{nongeodbyscale}. After changing variables to $r_I(x) = x+a_I(x)/2$ and likewise for $J$ and then relabelling them both $r$, we obtain:
\begin{align}
\frac{\Delta {\rm (length)}}{4G} & =
\int_{r=-a/2}^{r=a/2} S\left(dr \, \Big| \, \bar{a}_J(r) - dr\right)
-  \int_{r=-a/2}^{r=a/2} S\left(dr \, \Big| \, \bar{a}_I(r) -dr\right)
\nonumber \\ & =
\int_{-a/2}^{a/2} dr \left( \frac{dS(a)}{da} \Big|_{a = \bar{a}_J(r)} -
\frac{dS(a)}{da} \Big|_{a = \bar{a}_I(r)} \right) \geq 0
,\end{align}
where $\bar{a}_I(r_I(x)) = a_I(x)$ and likewise for $J$. Reading off $\bar{a}_{I,J}(r)$ is illustrated in Fig.~\ref{abars}; note that $\bar{a}_I(r) = r + a/2$. The inequality follows from the concavity of entropy since $\bar{a}_J(r) \leq \bar{a}_I(r)$.

To make this argument, we have implicitly assumed that the incremental merging protocol described here -- one whose entanglement cost is $\sdiff$ -- is the most efficient one possible given the constraints. We have sketched how to achieve the cost $\sdiff$, but we have not yet proved its optimality. Doing so is the purpose of Sec.~\ref{optimality}. Assuming the result, however, we find a remarkable new addition to the holographic dictionary. The Ryu-Takayanagi formula states that the entropy of a boundary interval $I$ is the length of the shortest bulk curve starting and ending at the endpoints of $I$.\footnote{More generally, one uses the shortest bulk curve homologous to $I$.} Entropy can be interpreted as the minimal number of qubits required to compress a state~\cite{schumacher} and, therefore, the minimal entanglement cost required to teleport it. We have demonstrated that those two minimizations, over bulk curves and boundary teleportation procedures, are effectively equivalent. Non-minimal length convex curves define constrained boundary state merging tasks whose optimal entanglement costs are the lengths of the curves themselves.

\subsection{Orientation reversal and negative length -- the ``cost'' of purifying a state}
\label{purify}

In fact, formula (\ref{main}) computes the signed length of an oriented curve. A detailed explanation of this can be found in \cite{lampros}; see also \cite{robproof}. The orientation of the curve is natural from the viewpoint of the merging protocol: it is decided by the direction of the flow of information.

To understand this, return to the geodesic $g_I$ drawn in Fig.~\ref{picgeodesic}. In Sec.~\ref{scalebyscale} we considered a stepwise merging protocol, in which Bob constructs the state on an interval $I$ of size $a$ from successive pieces received from Alice. Bob starts with the UV sector and builds up to the scale $a$. But we can consider the opposite situation, in which Bob initially holds the state on $I^c$, the complement\footnote{This discussion applies directly to pure states of the CFT. When the state of the CFT is mixed, such as the thermal state dual to a black hole spacetime, $I^c$ must be replaced by the total system which purifies $I$. For example, if we describe the thermal state as the thermofield double state of the composite system CFT$\otimes\widetilde{\rm CFT}$, then Bob holds $I^c \cup \widetilde{\rm CFT}$.} of $I$. Now Alice will send Bob the information about the purifier system $I$. She will again do so piecewise, this time starting from the data about the largest scale $a$ that is inaccessible to Bob and zooming down to the UV.

In the unconstrained merging protocol, Bob will use his full knowledge of the previously received state to merge each incoming chunk of information. To track his progress, we can use the intervals $I(x)$ from Fig.~\ref{picgeodesic}a, except now starting from $x=0$ up to $x = a/2$. This is convenient for comparison with Sec.~\ref{scalebyscale}, because it corresponds to complementing Bob's state from the left endpoint of the interval $I$ at $x = -a/2$ to the right endpoint at $x = a/2$, i.e. in the direction of increasing $x$. (Of course, Alice could send Bob data about the interval $I$ starting from the right endpoint, in which case the $I(x)$ with $-a/2 \leq x \leq 0$ would be natural.) Overall, prior to step $x$ Bob holds the state on $I(x-dx)^c$ while after step $x$ he knows the state on $I(x)^c$. The cost of this step in the merging protocol is
\begin{equation}
S\big(I(x)^c\big) - S\big(I(x-dx)^c\big) =  S\big(I(x)\big) - S\big(I(x-dx)\big),
\end{equation}
where we assume that the global state is pure. Adding up the cost of all steps, we get:
\begin{equation}
\int_{x = 0}^{x=a/2} \Big( S\big(I(x)\big) - S\big(I(x-dx)\big)\Big)  = S\big(I(a/2)\big) - S\big(I(0)\big) = -S(I)
\label{minusgeodlength}
\end{equation}
The ``cost'' of purifying a given mixed state is apparently negative! What this means is that rather than requiring an investment of Bell pairs, state merging \emph{produces} Bell pairs as a side effect~\cite{naturepaper}.

For readers unfamiliar with state merging, another simple (in fact, trivial) example may again be helpful. Suppose that Alice and Bob share the state $\ket{\psi}_{AB} = \smfrac{1}{\sqrt{2}}( \ket{00}_{AB} + \ket{11}_{AB})$ and that Alice wishes to merge $A$ to Bob, who holds $B$. The objective is to prepare a state in Bob's laboratory identical to $\ket{\psi}_{AB}$. The key point in this case, as compared to the merging examples in Sec.~\ref{explnongeodesic}, is the absence of a third system $R$. In general, the Alice-Bob merging procedure is required to maintain all the correlations with outside systems like $R$. In the absence of such systems, there are no correlations to preserve. Therefore, a perfectly good merging protocol is for Bob to prepare the state $\ket{\psi'} = \smfrac{1}{\sqrt{2}}(\ket{00}_{A'B'} + \ket{11}_{A'B'})$ in his own laboratory, without any help from Alice.  The initial state $\ket{\psi}_{AB}$ is left untouched so at the end of the merging protocol Alice and Bob share 1 Bell pair. This is exactly as it should be: for $\ket{\psi}_{AB}$, we have $S(A|B)=-1$ so instead of consuming a Bell pair, Alice and Bob return one. In more realistic situations, the state initially shared between Alice and Bob will be mixed. In that case, negative cost merging amounts to entanglement distillation: the extraction of good Bell pairs from noisy entanglement using only local operations and classical communication~\cite{bennett1996mixed,devetak2005distillation}.

Can \eqref{minusgeodlength} be interpreted as a differential entropy? Above we switched the roles of $I(x)$ and $I(x)^c$. If we substitute the complementary intervals in the definition (\ref{main}), we obtain:
\begin{equation}
\int \Big(S\big(I(x)^c\big) - S\big(I(x)^c \cap I(x-dx)^c\big) \Big) =
\int \Big(S\big(I(x)\big) - S\big(I(x) \cup I(x-dx)\big) \Big) \equiv \sdiff
\label{main2}
\end{equation}
We take this to be the definition of the differential entropy under reversal of orientation \cite{roblast, robproof, lampros}. This extension is sensible and necessary. Recall that the intervals $I(x)$ are defined by the requirement that the geodesics subtending them are tangent to the bulk curve. But this definition is ambiguous: if a family $I(x)$ satisfies it, so does $I(x)^c$. This ambiguity is fixed by a choice of orientation on the curve. Augmenting the definition by (\ref{main2}) makes it covariant under orientation reversal, whose boundary counterpart is to take the complement of each set $I(x)$.

Does this amendment make the definition ambiguous? Yes, but only up to a sign. Given a curve $R = R(\tilde{x})$ that subtends a boundary interval $I$, select a family of intervals $I(x)$. We can now compute the length of the curve using formula (\ref{main}) or using formula (\ref{main2}). One computes the number of Bell pairs required by Bob to learn the state on $I$ starting from nothing while the other computes the number of Bell pairs that can be extracted as Bob purifies his initial state on $I^c$. They always give opposite answers, because
\begin{equation}
S\big(I(x)\big) - S\big(I(x) \cup I(x-dx)\big) = - \Big( S\big(\tilde{I}(x)\big) - S\big( \tilde{I}(x) \cap \tilde{I}(x+dx) \big) \Big)
\label{sumdiff}
\end{equation}
for the family of intervals $\tilde{I}(x) = I(x) \cup I(x-dx)$, which is equivalent to $I(x)$ when $dx \to 0$.

\subsection{Closed curves: constrained state swapping}
\label{swapprotocol}
The preceding subsections consider the length of a convex curve with endpoints on the boundary. Its information theoretic interpretation involves sending the quantum state on the boundary interval lying between the curve's endpoints, subject to constraints that specify the shape of the curve. We now give a similar interpretation of the length of a closed, convex curve in AdS$_3$. This introduces several important differences.

\begin{figure}[t!]
\centering
\begin{tabular}{lp{1.5cm}r}
\raisebox{5.5cm}{a)}\includegraphics[width=.40\textwidth]{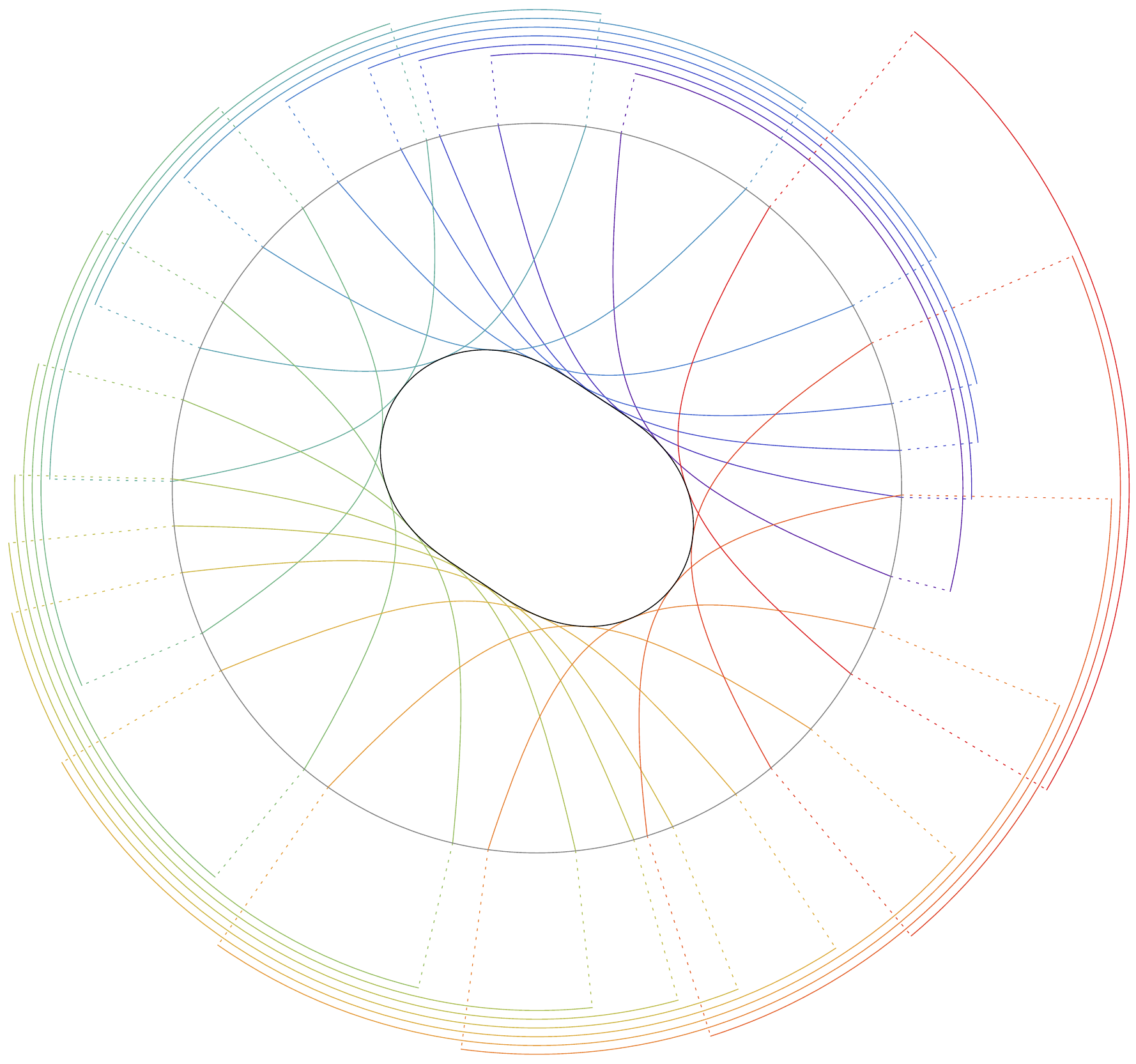}
&&
\raisebox{5.5cm}{b)}\includegraphics[width=.40\textwidth]{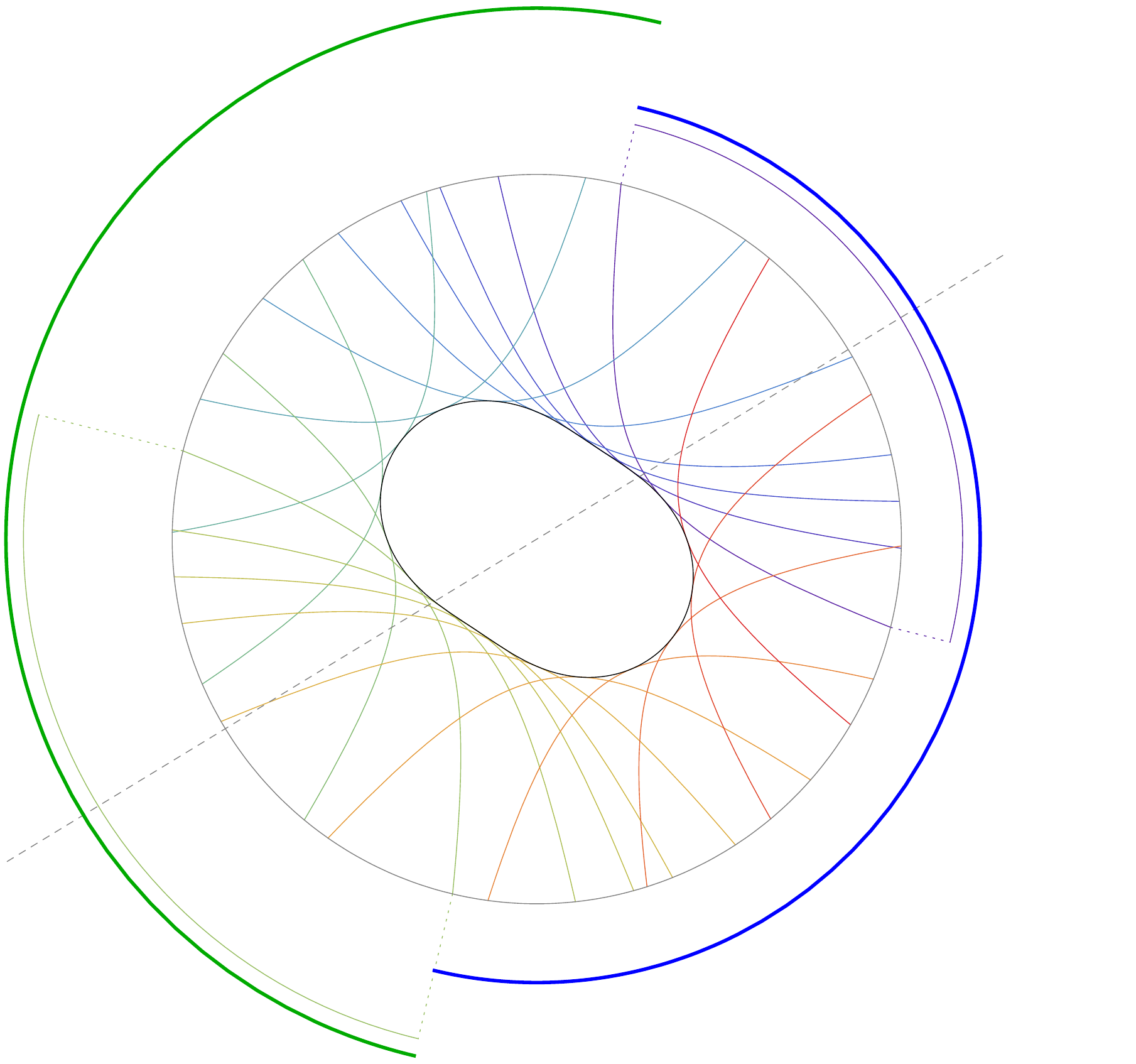}
\end{tabular}
\caption{a) The intervals $J(\theta)$ defined by a closed, convex curve in global AdS$_3$. b) The intervals (\ref{alice0}-\ref{bob0}) initially held by Alice and Bob. We have indicated $J(\theta_N)$ and $J(\theta_{2N})$ and the bulk axis, which joins their centers.}
\label{picswap}
\end{figure}

First, we can only speak of closed convex curves in global AdS$_3$ and not on the Poincar{\'e} patch. The dual field theory now lives on a circle instead of a line. Our curves will be given by $R = R(\tilde\theta)$ in coordinates
\begin{equation}
ds^2 = - \frac{R^2+L^2}{L^2}\, dT^2 + \frac{L^2}{R^2+L^2}\, dR^2 + R^2 d\tilde{\theta}^2
\label{ads3global}
\end{equation}
instead of $R(\tilde{x})$ in coordinates (\ref{ads3poincare}). Coordinate $\tilde\theta$ is an angular coordinate with period $2\pi$. As before, we distinguish the bulk coordinate $\tilde\theta$ from $\theta$, which we reserve for the asymptotic boundary. Following our earlier prescription, every point $\tilde\theta$ on $R = R(\tilde\theta)$ determines an interval $J(\theta)$ with center at $\theta$ and width $a_J(\theta)$, such that the geodesic subtending it is tangent to the curve. Once more, we caution that $\theta$ depends on $\tilde\theta$, but is not equal to it. The construction is illustrated in Fig.~\ref{picswap}a.

The second difference is that in contrast to the curves asymptoting to the boundary that we considered before, a closed bulk curve does not select a boundary interval. In consequence, the length of a closed curve does not compute the cost of sending the state on any one interval. Instead, we now consider Alice and Bob, who control complementary intervals on the CFT and wish to swap them. As before, the intervals $J(\theta)$ define a set of locality constraints, which limit the type of operations Alice and Bob can perform. The length of the closed curve is the total cost in Bell pairs for Alice and Bob to swap their states completely.

Our protocol will be carried out in discrete steps. To index them, we define $\theta_j = \pi j / N$ for $j = 1, \ldots, 2N$. Initially, let Alice and Bob control the states on:
\begin{eqnarray}
{\rm Alice:} & & \cup_{j = 1}^N J(\theta_j) - J(\theta_{2N}) \label{alice0} \\
{\rm Bob:} & & \cup_{j = N+1}^{2N} J(\theta_j) - J(\theta_{N}) \label{bob0}
\end{eqnarray}
A simple calculation in set arithmetic confirms that these intervals are complementary provided that $J(\theta_N) \cap J(\theta_{2N}) = \emptyset$. In pure AdS$_3$, this and the more general condition $J(\theta_j) \cap J(\theta_{j+N}) = \emptyset$ follows directly from the concavity of the bulk curve. As a consequence, note that Alice controls all of $J(\theta_N)$ and Bob controls all of $J(\theta_{2N})$. The intervals (\ref{alice0}-\ref{bob0}) are illustrated in Fig.~\ref{picswap}b.

Each discrete step consists of two parts. First, Alice sends to Bob the state on an infinitesimal piece on one end of her interval, i.e. $A_1=J(\theta_1)-J(\theta_{2N})$. Then Bob sends to Alice the state on an infinitesimal interval on the other end, $A_{N+1}=J(\theta_{N+1})-J(\theta_N)$. Both state transfers happen via constrained state merging. This means that Bob (Alice) can only use the operations on $B_1=J(\theta_1)\cap J(\theta_{2N})$ (respectively $B_{N+1} = J(\theta_{N+1}) \cap J(\theta_N)$) to merge the quantum state on $A_1$ (respectively $A_{N+1}$). For the Bob$\to$Alice transfer, we assume that $J(\theta_1) \cap J(\theta_N) = \emptyset$, so that Alice can utilize all operations on $J(\theta_N)$ to decode the message from Bob. At sufficiently large $N$ this assumption is true for every convex curve of finite size. At the end of this first step, the states controled by Alice and Bob are:
\begin{eqnarray}
{\rm Alice:} & & \cup_{j = 2}^{N+1} J(\theta_j) - J(\theta_{1}) \label{alice1} \\
{\rm Bob:} & & \cup_{j = N+2}^{2N+1} J(\theta_j) - J(\theta_{N+1})\label{bob1}.
\end{eqnarray}
We have used the periodicity in $\theta$ to rewrite $\theta_1 = \theta_{2N+1}$ in (\ref{bob1}).
Comparing with (\ref{alice0}-\ref{bob0}), we see that these intervals are of the same form as before, except the indices that set the interfaces between Alice and Bob have shifted by 1. Since both parts of the first step were constrained state merging, Alice and Bob have paid an entanglement cost equal to
\begin{equation}
S(A_1 | B_1) + S(A_{N+1} | B_{N+1})\,.
\end{equation}
To effect a full swap, we must shift the index of the interface by $N$, so we repeat the steps outlined above $N$ times. The total cost is the differential entropy $\sum_{j=1}^{2N} S(A_j | B_j)$.

\section{Optimality: Minimization over all possible constrained merging strategies} \label{optimality}

\subsection{The need for an optimality proof}

The goal of this article is to give an information theoretic interpretation of the length of a convex spacelike curve. So far, what we have demonstrated is a boundary merging procedure meeting a set of locality and scale constraints whose entanglement cost is the length of the curve. Who is to say, however, that we should pay any attention to the cost of that specific procedure? If we are to claim that the length of the curve is the cost of merging Alice's interval to Bob subject to the constraints, we need to be sure that there is not some other way of achieving the same goal but at a reduced entanglement cost.

This is a crucial point. Consider the special case of the interpretation of the entropy $S(I)$ as the logarithm of the effective Hilbert space dimension, that is, the number of qubits required to compress $I$.\footnote{Technically, it is $\Hmax^\epsilon(I)$ that is the effective Hilbert space dimension, but we will see in Sec.~\ref{sec:single-shot} that the two entropies are essentially interchangeable for a large $c$ CFT.} There are two halves to the interpretation. First, that there exists a subspace of dimension $2^{S(I) + \text{subleading}}$ containing nearly all the support of the density operator and, second, that no significantly smaller subspace can do so. Indeed, if there were a subspace of dimension $2^{S(I)/2}$ containing nearly all the support of the density operator, the effective Hilbert space dimension would obviously be at most $2^{S(I)/2}$, not $2^{S(I)}$.

In keeping with this credo, the purpose of this section is to complete the interpretation of the length of a curve by proving that no constrained merging procedure can have entanglement cost less than $\sdiff$.

Throughout this section we will be discussing the properties of general constrained merging protocols. In order not to create confusion, we will refer to the constrained merging protocol described in \secref{explnongeodesic} as the \emph{greedy constrained merging protocol} because at each stage as much state as possible is merged from Alice to Bob.

\subsection{Formal definition of constrained merging and statement of the theorem}

First we need to formally define the permissible procedures. Write $\mathcal{Q}_E$ for the Hilbert space $(\mathbb{C}^2)^{\otimes E}$ and let $\mathcal{H}^A \equiv \mathcal{H}^B \equiv \otimes_{x \in \mathbb{Z}} \mathcal{H}_x$, with $\dim \mathcal{H}_x < \infty$ and only a finite number of $x$ such that  $\dim \mathcal{H}_x > 1$ . Let $I \subseteq \mathbb{Z}$ be a finite interval and $I(x) \subseteq I$ itself be an interval for each $x$, such that the left and right endpoints $\ell(x)$ and $r(x)$ of the intervals are non-decreasing with $x$. Without loss of generality, let $I = [1,N] = \{1, 2, \ldots, N \}$. An \emph{$E$-ebit constrained merging protocol} consists of an $N$ step procedure. $\mathcal{H}^B$ is initially prepared in the fixed product state $\ket{00\cdots 0}_B$. Write $\mathcal{D}(\mathcal{H})$ for the density operators on $\mathcal{H}$. Step $x$ consists of an $A\leftrightarrow B$ LOCC transformation
\begin{equation}
\mathcal{D}( \mathcal{H}^A_{I(x)} \otimes \mathcal{H}_{I(x)}^B
	\otimes \mathcal{Q}_{E_x}^{A'} \otimes \mathcal{Q}_{E_x}^{B'} )
\longrightarrow
\mathcal{D}( \mathcal{H}^A_{I(x)} \otimes \mathcal{H}_{I(x)}^B
	\otimes \mathcal{Q}_{F_x}^{A'} \otimes \mathcal{Q}_{F_x}^{B'} )
\end{equation}
in which the space $\mathcal{Q}_{E_x}^A \otimes \mathcal{Q}_{E_x}^B$ is initially prepared as a maximally entangled state, with $\sum_{x \in I} (E_x - F_x)= E$. (The spaces $\mathcal{Q}_{F_x}^{A'} \otimes \mathcal{Q}_{F_x}^{B'}$ will be used to store entanglement distilled by the merging procedure in the event that the cost is negative.)

LOCC stands for Local Operations and Classical Communication.
The details of the definition of LOCC are a bit complicated~\cite{bennett1996mixed} but for the purposes of the optimality proof, it suffices to know that any LOCC transformation of a density operator $\rho_{AB}$ will have the form $\rho_{AB} \mapsto \sum_k F_k \otimes G_k \rho_{AB} F_k^\dagger \otimes G_k^\dagger =: \sum_k p_k \sigma^{(k)}_{AB}$. The index $k$ can roughly be thought of as recording the outcomes of the measurements that were part of the procedure. LOCC maps obey the inequality $S(\rho_B) \geq \sum_k p_k S( \sigma_B^{(k)} )$~\cite{bennett1996mixed}. That is, they cannot cause the entanglement entropy to increase on average. (They can, however, increase the entanglement for individual measurement outcomes $k$.)

Given an initial state $\ket{\psi}_A \in \mathcal{H}^A$, an $E$-ebit constrained merging protocol will produce an ensemble $( p_k, \sigma_{AB}^{(k)} )_k$ of final states. The protocol is said to have \emph{merging error $\epsilon$} if for all $\ket{\psi}$
\begin{equation} \label{eqn:merging-error}
\sum_k p_k \left\|
 \sigma_B^{(k)} - \proj{\psi}_B
\right\|_1 \leq \epsilon.
\end{equation}
While this is an operationally sensible definition, it turns out that just requiring the merging error to be small is not quite enough to ensure the optimality of $\sdiff$. Instead we will impose a slightly stronger condition, namely that at each of the $N$ steps of the protocol, those sites of $A$ that have never been acted upon, combined with those sites of $B$ that will never be acted upon again, are consistent with $\ket{\psi}$. This is a reasonable demand: at any given time the interval $I$ can be divided into three sections: the completed section, a portion under construction, and an untouched section. We will require that the completed and untouched sections be properly correlated. Formalizing that notion requires some further notation.

\begin{figure}[t!]
\centering
\includegraphics[width=.70\textwidth]{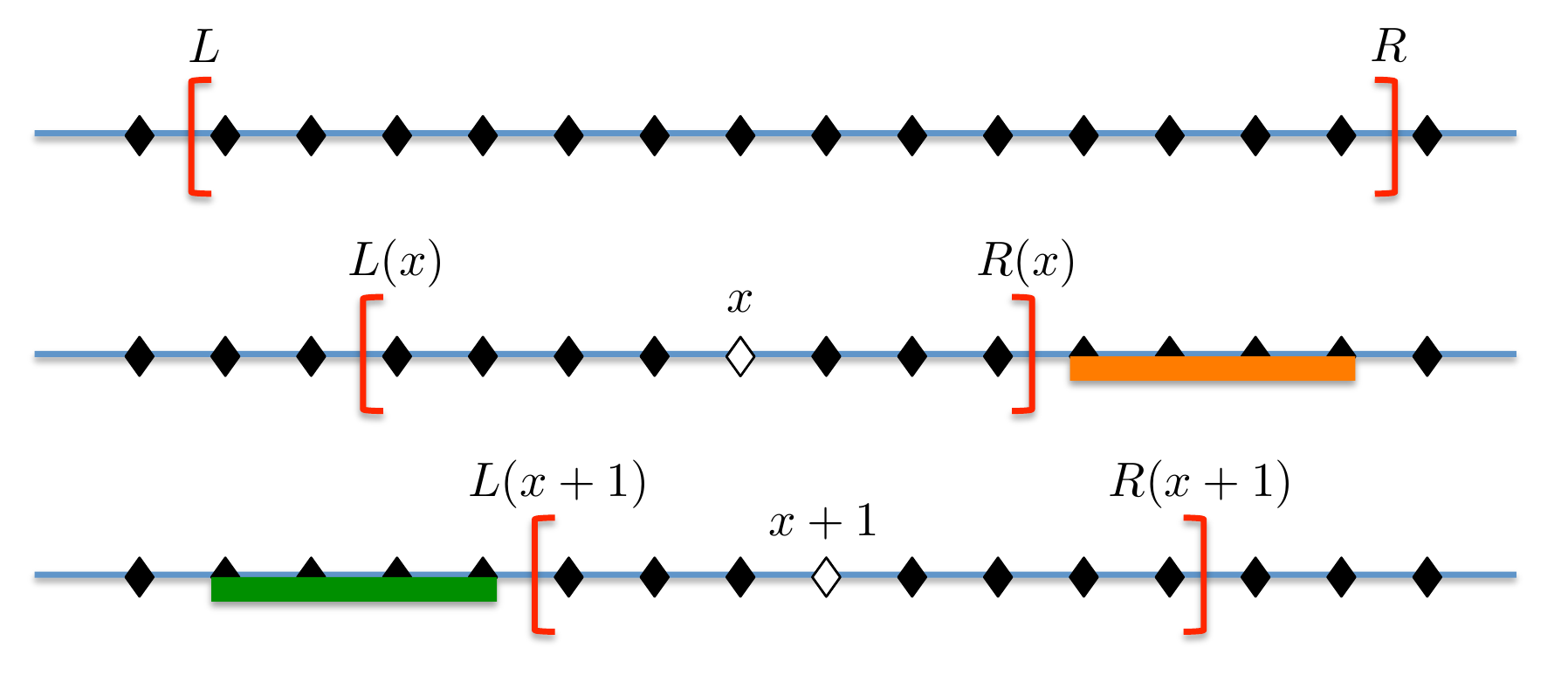}
\caption{Intermediate stage in a constrained merging protocol. The top row depicts the interval $I = [L,R]$ with endpoints $L$ and $R$. Step $x$ of the protocol acts on interval $I(x) = [\ell(x),r(x)]$, drawn in the second row. Because $r(x)$ is non-decreasing with $x$, the sites marked by the orange bar have not yet been acted upon. The third row depicts the interval $I(x+1) = [\ell(x+1),r(x+1)]$. Once step $x$ is complete, none of the sites indicated by the green bar will ever be acted upon again since $\ell(x)$ is non-decreasing with $x$. Therefore, after step $x$ the reduced density operators corresponding to the green marked sites on $B$ and the orange marked sites on $A$ should approximate the reduced density operator of the target state $\ket{\psi}$.}
\label{intervals}
\end{figure}

Let $I = \{ L, L+1, \ldots, R \} =: [L,R]$ so that $L$ and $R$ are the left and right endpoints of $I$. Likewise, let $I(x) = [\ell(x), r(x)]$. (We will also have occasion to make use of  abbreviations like $[L,\ell(x+1))^B$ to indicate the subsystem corresponding to $\ox_{x \in [L,\ell(x+1))} \mathcal{H}_x^B$.) After the step of the protocol acting on $I(x)$, the interval $(r(x),R] = \{ r(x) + 1, r(x) + 2, \ldots, R \}$ remains untouched in both $A$ and $B$. Similarly, the definition of constrained merging implies that none of the remaining steps $x+1,\ldots,N$ will act on the $[L,\ell(x+1))$ subsystem of $B$.  See \figref{intervals} for a visual depiction of these assertions.

\begin{figure}[t!]
\centering
\includegraphics[width=.70\textwidth]{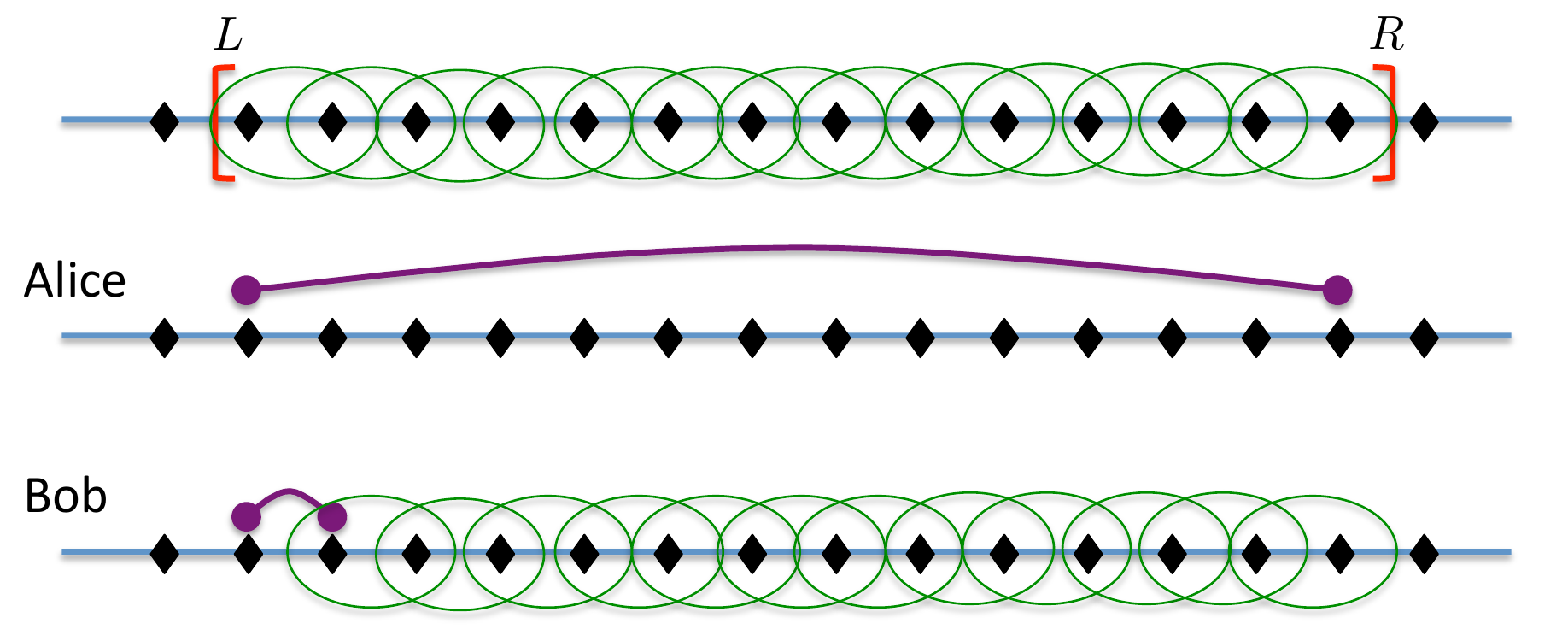}
\caption{Sequential merging error. The top line depicts the interval $I = [L,R]$ and a set of merging constraints in the form of intervals $I(x)$ depicted as green ovals. The state to be merged is a maximally entangled state shared between $L$ and $R$, as illustrated in the second line. One way to merge this state is for Bob to prepare a maximally entangled state between $L$ and $L+1$ in the first sequential merging step and then, in the subsequent steps, simply swap the information to the right until a maximally entangled state is established between $L$ and $R$. This procedure doesn't require any Alice-Bob entanglement and yet produces long-range entanglement in Bob's lab. It is prohibited by the sequential merging error condition because, while the final state is correct, the intermediate state fails the sequential merging condition. The joint state of Bob's $L$ and Alice's $R$ is not correct at the intermediate merging steps after $L$ has been completed and before $R$ has been reached; they should be maximally entangled but instead they are product.}
\label{sequential}
\end{figure}

Write $(p_{k,x},\sigma^{(k,x)}_{AB})_k$ for the ensemble of states produced after completion of the $x$ step of the protocol. Given the notational complexity, let us begin with the special case in which the protocol never generates entanglement so that $F_x = 1$ for all $x$.
An $E$-ebit constrained merging protocol is then said to have \emph{sequential merging error $\epsilon$} if for all initial states $\ket{\psi}\in \mathcal{H}^A$ and $x \in I$,
\begin{equation} \label{eqn:seq-merging-error0}
\sum_k p_{k,x} \left\|
 \operatorname{id}_{[L,\ell(x+1))}^{B \rightarrow A} \sigma^{(k,x)}_{[L,\ell(x+1))^{BA'B'} \cup (r(x),R]^A \cup (I^c)^A } -  \proj{\psi}_{\mathbb{Z} \setminus [\ell(x+1),r(x)]^A}
 \right\|_1 \leq \epsilon.
\end{equation}
The definition is quite subtle. Intuitively, it enforces the requirement that long-range entanglement in $\ket{\psi}$ be transferred from Alice to Bob rather than just manufactured entirely in Bob's laboratory, as illustrated in \figref{sequential}. The analogous conditions for $[L,\ell(x+1))^B$ and $(r(x),R]^A$ alone are in fact already consequences of the weaker definition \eqref{eqn:merging-error}. Small sequential merging error imposes the additional requirement that the joint density operator of $[L,\ell(x+1))^B$ and $(r(x),R]^A$, along with $I^c$, have the proper form. This ensures that the correlations Bob arranges in his lab are with the $(r(x),R]^A$ sitting in Alice's lab, not some new state he will manufacture himself later on in the protocol.

Requiring \eqref{eqn:seq-merging-error0} still leaves a great deal of freedom. In the case of a geodesic, for example, it is flexible enough to be consistent with both the all-at-once strategy of \secref{explgeodesic} and the greedy merging scale-by-scale strategy of \secref{scalebyscale}. In addition, after step $x$, the sites $[\ell(x+1),r(x)]^B$ are permitted be in arbitrarily messy intermediate states very different from the good approximations to $\psi$ produced at that stage by either of those merging protocols.

However, we will demonstrate that there is no advantage to be gained from all this freedom: greedy constrained merging is optimal.

The definition in the case in which the protocol can generate entanglement requires replacing \eqref{eqn:seq-merging-error0} with
\begin{equation} \label{eqn:seq-merging-error}
\sum_k p_{k,x} \left\|
 \operatorname{id}_{[L,\ell(x+1))}^{B \rightarrow A} \sigma^{(k,x)}_{[L,\ell(x+1))^{BA'B'} \cup (r(x),R]^A \cup (I^c)^A } -  \proj{\psi}_{\mathbb{Z} \setminus [\ell(x+1),r(x)]^A}
 					\otimes \proj{\Phi}_{[L,\ell(x+1))^{A'B'}}
\right\|_1 \leq \epsilon.
\end{equation}
Here, $\ket{\Phi}_{[L,\ell(x+1))^{A'B'}}$ is a product over $x \in [L,\ell(x+1))$ of maximally entangled states $\sum_{j=1}^{F_x} \ket{jj}_{A'B'}$. Thus, once the protocol has forever finished acting on some sites, the $A'B'$ portion of each of those sites must be left with high fidelity maximally entangled states of Schmidt rank $F_x$.
\begin{theorem}
For any $E$-ebit constrained merging protocol with sequential merging error $\epsilon < 1/4$, the following inequality holds for every initial state $\ket{\psi}$:
\begin{equation}
E \geq \sum_{x \in I} \left[ S\left( I(x) \right)_\psi - S\left( I(x) \cap I(x-1) \right)_\psi \right] - f( \epsilon ) \sum_{x \in I} \log \dim \mathcal{H}_x,
\end{equation}
where $f(\epsilon)$ vanishes as $\epsilon \rightarrow 0$.
\end{theorem}

\subsection{Proof}

To prove the theorem, begin by fixing an initial state $\ket{\psi}$ and an arbitrary $E$-ebit constrained merging protocol with sequential merging error $\epsilon$.  Again to keep notation relatively simple, we will assume that the protocol never generates any entanglement ($F_x=1$ for all $x$), the general case being a straightforward if cumbersome modification. We will write $\bar{S}(J)^B_x$ for the entropy averaged over reduced states on subsystem $J$ of $B$ that are produced after the step of the constrained merging protocol acting on interval $I(x)$.

There is an entanglement gain $\bar{S}(I(x))^B_x - \bar{S}(I(x))^B_{x-1}$ for the $I(x)$ step of the protocol, between the $I(x)$ portion of $B$ and its purification. Since the average entanglement entropy cannot increase under LOCC, we must have
\begin{equation} \label{entang-gain}
E_x  \geq \bar{S}(I(x))^B_x - \bar{S}(I(x))^B_{x-1}
\end{equation}
and, therefore,
\begin{align}
E
= \sum_{x \in I} E_x
&\geq \sum_{x \in I} \left[ \bar{S}(I(x))^B_x - \bar{S}(I(x))^B_{x-1} \right] \\
&= \sum_{x \in I} \left[ \bar{S}( I(x) )^B_x - \bar{S}( I(x+1) )^B_{x} \right],
\end{align}
where we have set $\bar{S}(J)^B_0 = 0$ to reflect that the initial $B$ state is a pure product state and  have defined $I(N+1)=\emptyset$.
As discussed above, after step $x$, the protocol has not yet acted on $(r(x),R]$ so $\sbr ( I(x+1) )^B_x = \sbr ( [ \ell(x+1), r(x) ] )^B_x$. That allows us to write
\begin{equation}
\bar{S}( I(x) )^B_x - \bar{S}( I(x+1) )^B_{x}
= \sbr\left( [ \ell(x), \ell(x+1) ) \big| [ \ell(x+1), r(x) ] \right)^B_x
\end{equation}
where, as usual, $\sbr (J|K)^B_x \equiv \sbr (JK)^B_x - \sbr (K)^B_x$.

By the sequential merging condition, $\sum_k p_{k,x} \epsilon_{k_x} \leq \epsilon$, where
\begin{equation}
\epsilon_{k,x} = \left\|
 \operatorname{id}_{[L,\ell(x+1))}^{B \rightarrow A} \sigma^{(k,x)}_{[L,\ell(x+1))^B \cup (r(x),R]^A \cup (I^c)^A }-  \proj{\psi}_{\mathbb{Z} \setminus [\ell(x+1),r(x)]^A}
\right\|_1.
\end{equation}
Since these two density operators are close to each other, their purifications are related by isometric Hilbert space transformations.
The Hilbert space purifying the $\psi$ mixed state is obviously $\mathcal{H}^A_{[\ell(x+1),r(x)]}$. The Hilbert space purifying the $\sigma^{(k,x)}$ mixed state, on the other hand, is $\mathcal{H}_{[L,r(x)]}^A \otimes \mathcal{H}_{\mathbb{Z} \setminus [L,\ell(x+1))}^B$. But outside of $[L,r(x)]$, the state $\sigma^{(k,x)}$ is just $\ket{00\cdots 0}$ so the purification can be taken to be a state of $\mathcal{H}_{[L,r(x)]}^A \otimes \mathcal{H}_{[\ell(x+1),r(x)]}^B$.
Using Uhlmann's theorem~\cite{uhlmann1976transition} and standard inequalities relating the trace distance and fidelity~\cite{fuchs}, we conclude that there exists an an isometry taking $\mathcal{H}^A_{ [\ell(x+1),r(x)]}$ to $\mathcal{H}^A_{[L,r(x)]}  \otimes  \mathcal{H}^B_{[\ell(x+1),r(x)]}$ that maps $\ket{\psi}$ to a state $f_1(\epsilon_{k,x})$-close to $\sigma^{(k,x)}$ where $f_1(\epsilon_{k,x})$ vanishes with $\epsilon_{k,x}$.

But then
\begin{align}
&\, S \left( [\ell(x), \ell(x+1) ) \big| [\ell(x+1),r(x)] \right)_\psi \nonumber \\
&\leq
S \left( [\ell(x), \ell(x+1) )^B \big| [L,r(x)]^A  \cup [\ell(x+1),r(x)]^B \right)_{\sigma_{k,x}}
	+ f(\epsilon_{k,x}) \log \dim \mathcal{H}_{[\ell(x),\ell(x+1))^B} \nonumber \\
&\leq
S \left( [\ell(x), \ell(x+1) )^B \big| [\ell(x+1),r(x)]^B \right)_{\sigma_{k,x}} + f(\epsilon_{k,x}) \log \dim \mathcal{H}_{[\ell(x),\ell(x+1))^B}. \label{isometry-ineq}
\end{align}
The function $f(\epsilon_{k,x})$ also vanishes with $\epsilon_{k,x}$.
The first inequality holds thanks to the existence of the isometry relating the two systems being conditioned upon and the Alicki-Fannes conditional entropy continuity inequality~\cite{alicki2004continuity}, while the second inequality is an application of strong subadditivity.

The function $f$ is concave and, for $\epsilon < 1/4$ also monotone, so
\begin{equation}
\sum_k p_{k,x} f( \epsilon_{k,x} ) \leq \sum_k f( p_{k_x} \epsilon_{k,x} ) \leq f ( \epsilon ).
\end{equation}
Averaging \eqref{isometry-ineq} over $k$ then summing over $x$ finally gives
\begin{align}
&\, \sum_{x \in I} \sbr\left( [ \ell(x), \ell(x+1) ) \big| [ \ell(x+1), r(x) ] \right)^B_x \nonumber \\
&\geq
\sum_{x \in I} S \left( [\ell(x), \ell(x+1) ) \big| [\ell(x+1),r(x)] \right)_\psi - f(\epsilon) \sum_{x \in I}\log \dim \mathcal{H}_x \nonumber \\
&= \sum_{x \in I} \left[ S( I(x) )_\psi - S( I(x) \cap I(x + 1) )_\psi \right] - f(\epsilon) \sum_{x \in I}\log \dim \mathcal{H}_x  \nonumber \\
&= \sum_{x \in I} \left[ S( I(x) )_\psi - S( I(x) \cap I(x - 1) )_\psi \right] - f(\epsilon) \sum_{x \in I}\log \dim \mathcal{H}_x ,
\end{align}
which completes the proof since we saw earlier that the first line was a lower bound on the entanglement cost $E$. $\square$

A few remarks are in order. The careful reader will have noticed that the theorem as stated only applies to systems with finite dimensional constituent Hilbert spaces $\mathcal{H}_x$. That is manifestly not true for the Hilbert space of the CFT, even after imposing the UV cut-off  $\mu$. Imposing the cut-off does, however, ensure that each of the entropies $S(x)$ is finite. In the limit of large central charge, therefore, we could compress the initial state $\ket{\psi}$ on each of the lattice sites $x \in I$, disturbing the state by a total amount $\epsilon'$. The compressed state would sit inside a Hilbert space satisfying the hypotheses of the theorem. Moreover, any constrained merging protocol with $\epsilon$ error for the original state would have error at most $\epsilon + \epsilon'$ for the compressed state by the triangle inequality. Therefore, $\sdiff$ is indeed a lower bound on the entanglement cost for constrained merging in the CFT.

\section{Single-shot versus von Neumann entropies} \label{sec:single-shot}

So as not to complicate the presentation in Sec.~\ref{diff-entropy}, we ignored two important issues. First, in the preceding discussion the conditional von Neumann entropy was identified as the entanglement cost in each of the $N$ merging steps of the greedy constrained merging protocol, but the cost in the single-shot setting appropriate for us here is in fact the smooth conditional max-entropy $\Hmax^\epsilon(A_j|B_j)$~\cite{oneshot}. $\Hmax^\epsilon$ asymptotes to the von Neumann entropy $S(A|B)$ in the limit of many copies of a state, but our procedure is intended to act on a single copy of the CFT state, so we must work with the max-entropy. Second, other than in extremely simple cases, achieving the optimal merging cost requires allowing small imperfections in the final state. Tracking the accumulation of those imperfections through the multistep protocol will be important.

The definition of the smooth conditional max-entropy is somewhat complicated~\cite{oneshot,chain} and will not actually be necessary. We will only need the following two facts:\footnote{The statements in the literature have slightly different forms, because they are formulated in terms of the ``purified distance'' instead of the trace distance. We have performed conversions in \secref{smooth-ent-ineq} at the expense of poorer scaling with $\epsilon$.}
\begin{enumerate}
\item For any $\epsilon > 0$ and quantum state $\ket{\psi}_{ABR}$, there exists a quantum state merging protocol with entanglement cost
\begin{equation} \label{ub-single-shot}
\Hmax^{\epsilon^4/169}(A|B) + 4 \log\left(\frac{1}{\epsilon}\right) + 2 \log_2 13
\end{equation}
producing a state with density operator $\rho_{ABR}$ such that $\| \proj{\psi} - \rho \|_1 \leq \epsilon$~\cite{oneshot}. (In fact, if the cost is negative then the final joint state of $ABR$ together with the entanglement is $\epsilon$-close to $\proj{\psi}$ tensored with a perfect maximally entangled state.)
\item $\Hmax^\epsilon(A|B) \leq \Hmax^{\epsilon^2}(AB) - \Hmin^{\epsilon^2/4}(B) + \text{const}$~\cite{chain}.
\end{enumerate}
In light of these results, to get an upper bound on the entanglement cost of merging an interval of the CFT it suffices to bound the smooth $\Hmax^\epsilon(AB)$ from above and the smooth $\Hmin^\epsilon(B)$ from below. The unconditioned min- and max- entropies are much simpler to define and work with.

The smooth min- and max- entropies are defined as follows. Given a state $\rho$, consider the set $B(\rho,\epsilon)$ of all $\sigma$ with $\tr(\sigma)\leq 1$, $\sigma \geq 0 $, and $\|\sigma - \rho\|_1 < \epsilon$. This set is relevant because we want to determine the optimal resource cost, but without the unrealistic assumption that the state $\rho$ is perfectly transmitted. The smooth max-entropy $\Hmax^\epsilon$ is then defined as
\beq
\Hmax^\epsilon(\rho) = \min_{\|\sigma -\rho\|_1 < \epsilon} \log(\text{rank}(\sigma))\,,
\eeq
where the minimum is taken over $B(\rho,\epsilon)$. In essence $\Hmax^\epsilon$ instructs us to truncate $\rho$ to its largest eigenvalues of total weight $1-\epsilon$. In other words, we throw away the smallest eigenvalues of $\rho$ up to weight $\epsilon$, but then we must transmit the remaining state in its entirety since this is a single-shot protocol. Once more, we are allowed to ignore very rare events, but once these events have been cut out of $\rho$, all the remaining states must be sent to guarantee that the protocol succeeds in a single shot.

The smooth min-entropy $\Hmin^\epsilon$ is defined as
\beq
\Hmin^\epsilon(\rho) = \max_{\|\sigma-\rho\|_1 < \epsilon} \log\lamm^{-1}(\sigma)\,,
\eeq
where $\lamm(\sigma)$ is the largest eigenvalue of $\sigma$. For the smooth min-entropy we are doing a similar kind of truncation as with the smooth max-entropy, except that now we are truncating the largest eigenvalues of $\rho$ up to weight $\epsilon$. The smooth min-entropy is then the negative logarithm of the new largest eigenvalue after the truncation, which is a measure of one's ability to guess the quantum state correctly. As noted above, it plays an important role in bounding the single-shot state merging cost.

If the state $\rho$ were an equal weight mixture of $M$ pure states, it would immediately follow that both $\Hmax^\epsilon$ and $\Hmin^\epsilon$ are within $\epsilon$ of the von Neumann entropy $\log(M)$. In this case the single-shot cost is the same as the asymptotic cost (as it should be). What we now show is that, to leading order in the central charge $c$, the smooth min- and max- entropies for an interval in a CFT are also given by the interval's von Neumann entropy. Hence for CFT intervals (with some care taken about the errors) we find that the single-shot cost approximately reproduces the asymptotic cost.

To compute these smooth entropies, we use a formula found by Calabrese and Lefevre for the eigenvalue distribution of an interval's reduced density operator in either a vacuum or thermal state of a 1+1 dimensional CFT~\cite{calabrese-eigs}:
\beq
P(\lambda) = \sum_i \delta(\lambda - \lambda_i) = \delta(\lambda - \lamm) + \frac{b\, \Theta(\lamm - \lambda)}{\lambda \sqrt{b \log(\lamm/\lambda)}} I_1(2 \sqrt{b\log(\lamm/\lambda)})\,,
\eeq
where $b = \Hmin(I) = -\log(\lambda_{\max})$ and $I_1$ is a modified Bessel function of the first kind. This formula is derived starting from the fact that the R\'enyi entropies of an interval $I$, $S_\alpha(I) = \frac{1}{1-\alpha}\log(\tr(\rho_I))$, are given by
\beq \label{cft-renyi}
S_\alpha(I) = \frac{1}{2}\left(1 +\frac{1}{\alpha}\right) S(I),
\eeq
where $S(I)$ is the von Neumann entropy. The limit $\alpha \rightarrow \infty$ gives $S_\infty = \Hmin$, so $b = S(I)/2$. Furthermore, conformal invariance fixes $S(I)$ in the ground state to be
\beq
S(I) = \frac{c}{3}\log\left(\frac{a}{\mu}\right),
\eeq
where $a$ is the size of the interval $I$. Finally, it should be noted that for a given regulator, the form (\ref{cft-renyi}) will be corrected due to irrelevant operators, but for our purposes the important feature is that these corrections are expected to be subleading in $c$, the central charge.

Given that $b \propto c$, the large central charge limit corresponds to the limit of large $b$. Investigation of the limit $b \rightarrow \infty$ reveals the distribution of eigenvalues of $\rho_I$ to be well-approximated by a tightly peaked Gaussian after a suitable change of variables, yielding\begin{align} \label{eqn:smooths}
\Hmin^{\epsilon'}(I) &= S(I) \left( 1 - O\left( \sqrt{2  \log(1/\epsilon') / S(I)} \right) \right)
	\quad \text{and} \nonumber \\
\Hmax^{\epsilon'}(I) &= S(I) \left( 1 + O\left( \sqrt{2  \log( 1/\epsilon') / S(I)} \right) \right).
\end{align}

These estimates follow from a straightforward change of variables. Setting $y(\lambda) = 2 \sqrt{ b \log(\lamm/\lambda)} \geq 0$, the distribution of eigenvalues becomes
\beq
\tilde{P}(y) \equiv P(\lambda(y)) \left|\frac{d\lambda}{d y}\right| = I_1(y),
\eeq
where we have neglected the $\delta$ function at $\lamm$ (dropping one eigenvalue makes no difference at large $c$). Large $b$ corresponds to large $y$ where $I_1$ is well approximated by the form $I_1(y) \approx e^y/\sqrt{2 \pi y}$. Thus, the distribution of eigenvalues is roughly exponential in the $y$ variable. The leading correction to the asymptotic behavior is given by
\beq
I_1(y) = \frac{e^y}{\sqrt{2\pi y}}\left(1 - \frac{3}{8y} + O\left(\frac{1}{y^2}\right)\right),
\eeq
which is only a minor power-law correction.

The normalization $\sum_i \lambda_i$ translates to the statement that
\beq
\int_0^\infty dy \tilde{P}(y) \lambda(y) = 1,
\eeq
with $\lambda(y) = \lamm e^{-\frac{y^2}{4b}} = \lamm e^{-\frac{y^2}{2S}}$. The object $\tilde{P}(y) \lambda(y)$ at large $y$ is well approximated by
\beq
\tilde{P}(y) \lambda(y) \approx \frac{e^y}{\sqrt{2 \pi y}} \lamm e^{- \frac{y^2}{2S}} \approx \lamm \frac{e^{- \frac{(y-S)^2}{2S}}}{\sqrt{2 \pi S}}
\eeq
where in the last step we replaced $\sqrt{y}$ with $\sqrt{S}$ (its central value) in the denominator, a mild simplifying approximation. The final form of $\tilde{P}(y) \lambda(y)$ is thus a normalized Gaussian with polynomial in $(y-S)$ corrections (arising from subleading terms in $I_1$ and the expansion of $y^{-1/2}$ about $y=S$):
\beq
\tilde{P}(y)\lambda(y) = \lamm \frac{e^{- \frac{(y-S)^2}{2S}}}{\sqrt{2 \pi S}}\left(1 + O\left(\frac{y-S}{S}\right)\right).
\eeq
Once the tightly peaked nature of the Gaussian is taken into account, all subleading corrections are $O(S^{-1/2})$ or smaller.

To compute the smooth entropies with generic smoothing parameter $\epsilon'$, the probability distribution $\tilde{P}(y)\lambda(y)$ must be truncated to weight $1-\epsilon'$. Define $y_{\min,\max}$ by the equations:
\begin{eqnarray}
1-\epsilon' & = & \int_{y_{\min}}^\infty dy \frac{e^{-\frac{(y-S)^2}{2S}}}{\sqrt{2 \pi S}}\\
1-\epsilon' & = & \int_{-\infty}^{y_{\max}} dy \frac{e^{-\frac{(y-S)^2}{2S}}}{\sqrt{2 \pi S}}
\label{truncation2}
\end{eqnarray}
In (\ref{truncation2}) we assumed, up to corrections of order $e^{-S}$, that $y$ runs over the whole real line. It follows from symmetry that $y_{\min} = S - \delta y$ and $y_{\max} = S + \delta y$, where
\beq
\delta y(\epsilon') = \sqrt{2 S \log\left(\frac{1}{\epsilon'}\right)}+...
\eeq

The corrections may be bounded with elementary properties of the error function. With $\text{erf}(x)$ defined as
\beq
\text{erf}(x) = \frac{2}{\sqrt{\pi}}\int_0^x dt\, e^{-t^2}
\eeq
it follows that:
\beq
1-\epsilon' = \frac{1}{2} + \frac{1}{2}\text{erf}\left(\frac{\delta y}{\sqrt{2S}}\right).
\eeq
The asymptotic form of the error function is
\beq
\text{erf}(x) = 1 - \frac{e^{-x^2}}{x\sqrt{\pi}}\left(1 + O\left(\frac{1}{x^2}\right)\right),
\eeq
so to leading order
\beq
\frac{e^{-\frac{\delta y^2}{2S}}}{ \left(\frac{\delta y}{\sqrt{2S}}\right) 2\sqrt{\pi}} = \epsilon'.
\eeq
An upper bound on $\delta y$ is obtained by neglecting the denominator on the LHS, which gives the estimate:
\beq
\delta y \leq \sqrt{2 S \log\left(\frac{1}{\epsilon'}\right)}.
\eeq

The smooth max entropy is then the logarithm of the rank of the truncated state (the number of eigenvalues), yielding
\beq
\Hmax^{\epsilon'} = \log\left(\int_0^{y_{\max}} dy\, \tilde{P}(y)\right) \approx \log\left(\int^{y_{\max}} dy\, e^y \right) \approx y_{\max} = S + O(\delta y(\epsilon')).
\eeq
Similarly, the smooth entropy is the logarithm of the largest remaining eigenvalue after truncating the largest eigenvalues, so
\beq
\Hmin^{\epsilon'} = -\log(\lambda(y_{\min})) = - \log(\lamm) + \frac{y_{\min}^2}{2S} = \frac{S}{2} + \frac{(S - \delta y)^2}{2 S} \approx S - O(\delta y(\epsilon')).
\eeq
Pulling out a factor of $S$, the smooth entropies take the claimed form \eqref{eqn:smooths}.

We now have the tools to bound the parameters required to achieve a cumulative error of $\delta$ in an $N$-step constrained merging protocol. Each step should contribute an error of at most $\delta/N$, which means that $\epsilon' = \text{poly}(\delta/N)$ in \eqref{eqn:smooths} thanks to \eqref{ub-single-shot}. The error term in \eqref{eqn:smooths} is then
\begin{equation}
O\left( \sqrt{ \frac{\log N + \log 1/\delta}{c \log a_I/\mu} } \right),
\end{equation}
where $a_I$ is the length of the entire interval $I$. $N$ must be allowed to go to infinity but not so quickly as to invalidate the R\'enyi entropy formula \eqref{cft-renyi}. Choosing $N=(a_I/\mu)^\gamma$ for $0 < \gamma < 1$ is sufficient, in which case taking the limit of vanishing UV cut-off $\mu$ implies that the cost of each merging step is bounded above by the conditional von Neumann entropy, with a multiplicative correction of order $O(1/\sqrt{c})$. The total entanglement cost of the greedy constrained merging procedure is therefore precisely the differential entropy, up to the same $O(1/\sqrt{c})$ corrections.

\section{Differential entropy and Markov chains}
\label{secmarkov}

From the point of view of the boundary field theory, the differential entropy is an entropic function of a collection of reduced density matrices. In the first part of this work we provided an information theoretic interpretation for differential entropy in terms of the entanglement cost of a restricted communication task. However, there may be alternative interpretations in terms of the entanglement entropy of a reconstructed global state. In fact, it has been conjectured that the differential entropy is the maximum entropy among all global states consistent with the marginals~\cite{holeography}. (In the reconstructability literature, the reduced density matrices are referred to as marginals, a nomenclature we follow in this section.) Arguments against this conjecture were first given in \cite{veronikajune}. Recently, Kim and Swingle showed that this conjecture is false by arguing that it does not apply to a global  pure state or subsystems with local modular Hamiltonians \cite{swingle-kim}. While their argument disproves the conjecture, there remains an intriguing connection between differential entropy and the problem of reconstructing the global state. Indeed, the greedy constrained merging protocol is nothing but an operational way of reconstructing the global state by assembling its marginals. We believe that density matrix reconstructability deserves further investigation, as it may be relevant for thermal states or highly excited energy eigenstates \cite{swingle-kim}. Hence, we devote this section to the study of reconstructability and its intimate connection with quantum Markov chains.

\subsection{Reconstructability and Markov chains}

Given a set of marginals, there can be many global states consistent with that local data because local data need not fix long-range correlations.
In fact, it is unlikely that the set of all consistent global states can even be characterized efficiently \cite{poulin2011markov}.\footnote{Given a set of local density matrices, just verifying their consistency is already QMA-complete, the quantum computing version of NP-complete~\cite{liu2006consistency}. Here, however, we know in advance that local states are consistent.
} Nevertheless, all consistent global states should satisfy certain entropic inequalities. In particular, the entropy of a consistent global state cannot be arbitrarily large. In the absence of further information, the best guess for the global state is the state that maximizes global entropy; we denote this state by $\rho_{\max}$.

Suppose we are given a set of local density matrices $\{ \rho_{I_j} \}$ for the intervals $\{ I_j \}$. If the $I_j$ do not overlap, then the subadditivity of entropy requires that
\bea \label{subadd-marg}
S(\rho)\leq \sum_j S(\rho_{I_j})
\eea
for any global state $\rho$ consistent with the marginals.
The product state $\otimes_j \rho_{I_j}$ saturates the above inequality and is therefore equal to $\rho_{\max}$. \eqref{subadd-marg} also suggests a strategy for moving beyond this simple situation. The quantity $\sum_j S(\rho_{I_j}) - S(\rho)$, which quantifies ``unnecessary'' correlations in $\rho$, is nothing other than the relative entropy $S(\rho \| \rho_{\max} )$ provided $\rho$ is consistent. This relative entropy measure of deviation from the maximum entropy consistent state generalizes naturally to the case of overlapping intervals.

Reconstructing the global state from local data is much more interesting when regions overlap. The density matrix of two overlapping regions $A$ and $B$ always satisfies the strong subaddivity of entropy:
\bea\label{SSA}
0\leq I(A -B : B - A|A\cap B)\equiv S(A)+S(B)-S(A\cap B)-S(A\cup B),
\eea
where $I(A-B:B-A|A\cap B)$ is the conditional mutual information. \eqref{SSA} is saturated if and only if the quantum state decomposes into a quantum Markov chain $A-B \rightarrow A \cap B \rightarrow B - A$~\cite{hayden2004structure,fawzi2014quantum}. Informally, that means that there exists an incomplete projective measurement of $A \cap B$ leaving the state invariant but such that $A - B$ and $B-A$ factorize conditioned on the measurement outcome.  In other words, all correlations between $A-B$ and $B-A$ are classical correlations mediated by $A\cap B$.
(See Appendix \ref{ssa-saturation} for more precise statements.)

In analogy with the case of nonoverlapping intervals in which $\rho_{\max}$ was the state saturating \eqref{subadd-marg}, in the general case it would be natural to expect $\rho_{\max}$ to saturate \eqref{SSA}. Any consistent state saturating strong subadditivity is uniquely determined from the local data according to the prescription
 \bea\label{saturate}
 \log \rho_{A\cup B}=\log \rho_A+\log \rho_B-\log \rho_{A\cap B}.
 \eea
(Equality in \eqref{saturate} is in fact a necessary and sufficient condition for the saturation of strong subadditivity~\cite{ruskai}.) This equation gives a prescription for writing down $\rho_{A \cup B}$ that will always yield a positive semidefinite Hermitian operator. In general, however, $\rho_{A \cup B}$ need not have unit trace. When it does not, $\rho_{\max}$ will necessarily fail to saturate strong subadditivity.

The neighboring regions in the formula for differential entropy, $I(x)$ and $I(x-dx)$, have significant overlap. In this case, the conditional mutual information computed for a 2D CFT is negligible:
\bea
S(I(x))+S(I(x-dx))-S(I(x)\cap I(x-dx))-S(I(x)\cup I(x-dx))= O(dx^2).
\eea
Therefore, one might hope that the maximum entropy consistent global state could be found by a repeated application of (\ref{saturate}) using the local data. Assuming the saturation of strong subadditivity for all neighboring regions and iterating \eqref{saturate} over all $j$ defines an operator $\sigma$ satisfying
\bea\label{rho_global}
\log\sigma=\sum_j\lb \log\rho_{I_j}-\log\rho_{I_j\cap I_{j-1}}\rb,
\eea
which we call the Markov operator. In those cases when $\sigma$ is normalized, it defines a state consistent with the marginals whose only long-range correlations are purely Markovian.

A key observation is that if the global state is a pure state $|\Psi\ra$, then the formula for the differential entropy in (\ref{eqn:discretized}) is nothing but the relative entropy of $|\Psi\ra$ with respect to $\sigma$:
\bea
\sdiff=S\left(|\Psi\ra\la\Psi|\big\|\sigma\right)=S(\sigma),
\eea
as can be verified by a simple calculation.
If $\sigma$ were properly normalized, this would have constituted a proof for the conjecture that the differential entropy is the entropy of the maximum entropy consistent state.

In appendix \ref{golden} we demonstrate that on a line (without periodic boundary conditions) $\tr(\sigma)\leq 1$. This inequality is saturated in the special case where all $\rho_{I_j}$ and $\rho_{{I_j}\cap I_{j-1}}$ commute. Therefore, the Markov operator $\sigma$ in this case is the consistent global state with maximum entropy, and its von Neumann entropy is the differential entropy.

\subsection{Markov operator in conformal field theory}

In this section, we show explicitly that the Markov operator $\sigma$ corresponding to marginals of size $R$ in the vacuum of a 1+1-dimensional conformal field theory does not have unit trace. We thereby give a quantitative refutation of the conjecture that the entropy of the maximum entropy consistent state is $\sdiff$. We will show that the maximum entropy state consistent with marginals of size $R$ in the vacuum has entropy at most $2/3$ of $\sdiff$.

The reduced density matrix of a region $A$ of size $2R$ in a 1+1-dimensional conformal field theory on a line is \cite{casini2011towards}
\bea\label{marginals}
\rho_A=\exp\lb -2\pi\int dx \frac{R^2-x^2}{2R}\:T_{00}(x)\rb.
\eea

As we saw in the previous section, the sufficient condition for differential entropy to be the entropy of a consistent state is that reduced density matrices on different intervals commute. Naively, it appears that reduced density matrices should commute because they are functions of only one operator, $T_{00}$. However, it is well known that in field theory equal-time commutators of symmetry currents need not be zero \cite{schwinger1963commutation}. In particular, in relativistic field theories the so-called Schwinger term quantifies the amount by which stress tensors fail to commute:
\bea
i[T^{00}(x),T^{00}(y)]=\lb T^{0k}(x)+T^{0k}(y)\rb \partial_k \delta(x-y).
\eea

According to \eqref{rho_global}, knowledge of the reduced density matrices in (\ref{marginals}) is sufficient to directly compute the Markov operator. In \appref{cft-calc} we find $\sigma$ for a collection of marginals of size $R$ in a CFT on a line to be
\bea
\sigma=C\: \exp\lb-\int dx \frac{4\pi R}{3}\: T_{00}(x)\rb,
\eea
which is proportional to the thermal state at temperature $T=\frac{3}{4\pi R}$.
The identity $\sdiff=S(|\Omega\ra\la\Omega|\|\sigma)$ fixes the value of $C$ when the global state is the vacuum $|\Omega\ra$:
\bea
\sdiff=-\la\Omega|\log\sigma|\Omega\ra=-\log C+\frac{4\pi R}{3}\int dx\:\la\Omega| T_{00}|\Omega\ra=-\log C.
\eea
Therefore,
\bea
\sigma=\rho_T Z_T e^{-\sdiff}= \rho_T e^{\frac{c L}{6 R}-\frac{c L}{8 R}}=\rho_T e^{-\frac{c L}{24 R}},
\eea
where $\rho_T$ and $Z_T$ are the thermal state and partition function at temperature $T=\frac{3}{4\pi R}$, respectively. In this case, $Z_T=e^{S_{\operatorname{thermal}}/2}=e^{c L/ (8 R)}$, and $\sdiff=\frac{c L}{8 R}$.

Consider $\rho_{\max}$, the maximum entropy consistent density matrix for the set of marginals of size $R$. The relative entropy of this state with respect to $\sigma$ is
\bea
S(\rho_{\max}\|\sigma)=\sdiff-S(\rho_{\max})
\eea
by direct substitution of the definition of $\sigma$ into the formula for the relative entropy.
However, as we have seen, $\sigma=\rho_T e^{-\frac{c L}{24 R}}$. Using the nonnegativity of relative entropy then leads to
\bea
\sdiff-S(\rho_{\max})=S(\rho_{\max}\|\rho_T)+\frac{c L}{24 R}\geq \frac{c L}{24 R}.
\eea
Therefore, there exists no consistent density matrix with an entropy that is near the differential entropy; the largest entropy among all of them is at most 2/3 of $\sdiff$.

As a final remark, it is worth noting that the thermal density matrix at temperature $T=\frac{3}{4\pi R}$ is not consistent with vacuum at scales smaller than $2R$. In fact, the thermal state at temperature $T$ remains distinguishable from the vacuum in the large central charge limit since $S(\rho_T\|\|\Omega\ra\la\Omega|)=O(c)$.

\section{Generalization to more curves and surfaces}
\label{disc}
Our discussion of the constrained merging protocol was given mostly for a convex curve with endpoints on the boundary on a static slice of pure AdS$_3$, which is dual to the ground state of a 1+1-dimensional CFT. In fact, our interpretation of the length of a convex curve works for a broader class of curves and even extends to areas of some surfaces in higher dimensions.

\begin{figure}[t!]
\centering
\raisebox{4cm}{a)}\includegraphics[width=.45\textwidth]{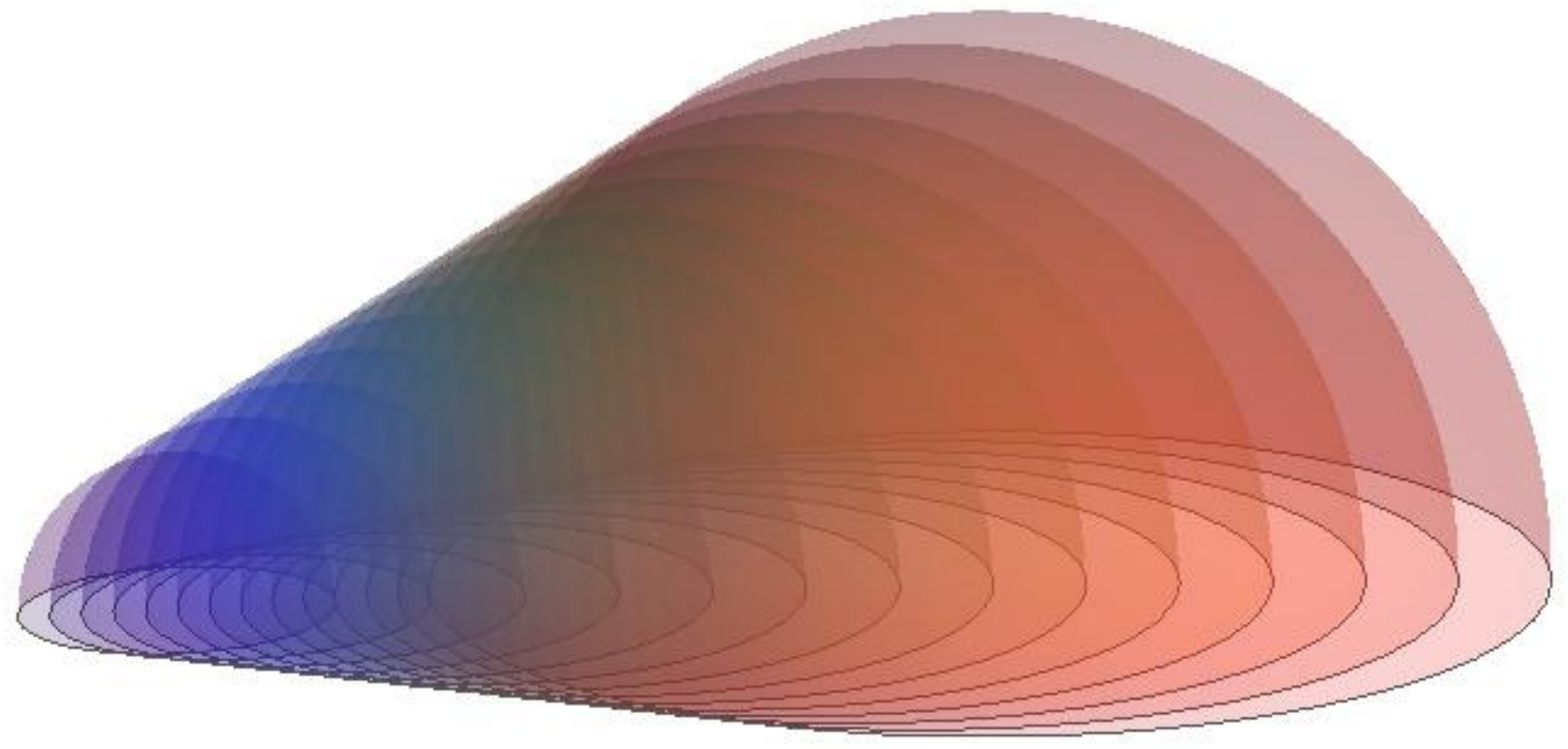}
\raisebox{4cm}{b)}\includegraphics[width=.45\textwidth]{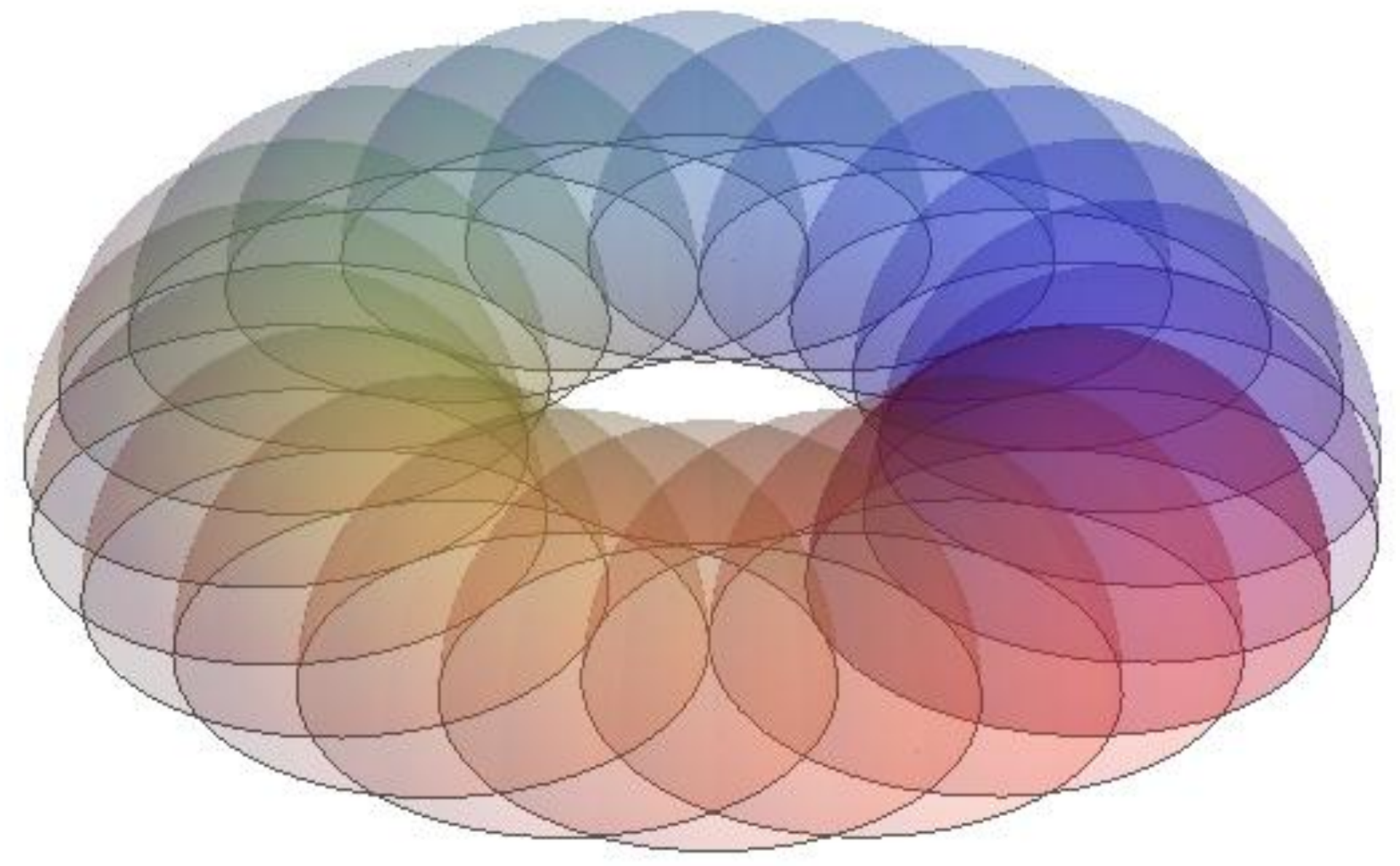}
\caption{a) A surface in Poincar{\'e} AdS$_4$ built up from minimal surfaces anchored on a sequence of boundary circles. Its area computes the entanglement cost of a constrained merging protocol. b) A surface built up from a cycle of minimal surfaces. Its area computes the entanglement cost of a constrained swapping protocol.}
\label{lastfig}
\end{figure}

A key geometric fact underlying the scope of our results is that the area of any spacelike convex surface in a holographic spacetime can be written in the form
\begin{equation}
{\rm area}_{\rm surface} =
\lim_{N \to \infty} \sum_{j = 1}^N ({\rm area}_{A_j\cup B_j} - {\rm area}_{B_j})
\label{genarea}
\end{equation}
for some family of sets $I_j$, where as before $A_j = I_j - I_{j-1}$ and $B_j = I_j \cap I_{j-1}$  \cite{xi, robproof}.
This includes lengths of curves in asymptotically AdS$_3$ spacetimes which do not lie on a constant time slice \cite{robproof}. The areas on the right hand side of (\ref{genarea}) are of extremal surfaces which asymptote to $A_j \cup B_j$ and $B_j$. So long as these areas compute appropriate boundary entanglement entropies, the total area on the left hand side becomes a sum of conditional entropies:
\begin{equation}
{\rm area}_{\rm surface} =
\sdiff = \lim_{N \to \infty} \sum_{j = 1}^N S(A_j | B_j)\, .\label{genform}
\end{equation}
For example, Fig.~\ref{lastfig} depicts two surfaces on a static slice of the Poincar{\'e} patch of AdS$_4$. Their areas compute the entanglement costs of the constrained merging and constrained swapping protocols discussed in Sec.~\ref{diff-entropy}.

However, applying our results in more general settings is subject to a number of caveats, which we discuss below.

\paragraph{Extremal but nonminimal surfaces}
The surfaces appearing on the right hand side of \eqref{genarea} are guaranteed to be extremal, but not necessarily minimal. In pure AdS spacetimes and some excited geometries \cite{Nogueira:2013if, Gentle:2013fma} this distinction does not play a role. In generic holographic spacetimes, however, more than one extremal surface may be anchored on the same boundary region \cite{renyis, Hubeny:2012ry, plateaux} and only one of them computes the entanglement entropy of the said region. In discussions of the Ryu-Takayanagi proposal, the non-uniqueness of the extremal surfaces came to focus with the introduction of the homology constraint, without which the proposal gives incorrect answers for entanglement entropies of subregions in the thermal state \cite{renyis}. Another easy example of an extremal but non-minimal surface is a geodesic in the BTZ spacetime, which wraps around the black hole multiple times in a spacelike analogue of gravitational lensing.

In order to interpret the area of a surface written in the form (\ref{genarea}) as the cost of a merging protocol (\ref{genform}), all surfaces on the right hand side must be minimal. It is difficult to give a general characterization of bulk surfaces for which this is true; for a discussion of this consult \cite{xi}. Here we content ourselves with some qualitative rules of thumb. The right hand side of \eqref{genarea} typically involves extremal but nonminimal surfaces (a) if the surface approaches a horizon, (b) if its extrinsic curvature is anywhere large, and / or (c) if it is locally  approximately radial. In addition, closed convex curves in higher dimensions always involve nonminimal surfaces in \eqref{genarea}.

It would be interesting to understand \eqref{genarea} in information theoretic terms also in cases, where nonminimal surfaces make an appearance. A possible starting point was given in \cite{entwinement}, which studied the field theory meaning of the lengths of nonminimal surfaces in the simplest setting: the conical defect geometry AdS$_3 / \mathbb{Z}_n$. The field theory state dual to this geometry is an excited state, so the level spacing in its neighborhood is reduced relative to the vacuum. Converting the level spacing into a length scale, we obtain a scale larger than system size -- a dynamical scale, which cannot be spatially realized in a single copy of the system. In a certain technical sense, \cite{entwinement} associated nonminimal geodesics with the physics of such extended, dynamical scales. Now recall that in Sec.~\ref{scalebyscale} we used differential entropy to decompose entanglement -- that is, a minimal surface -- into scale-specific components. This suggests that it may be possible to interpret \eqref{genarea} in information theoretic terms in the general case, perhaps in terms of a communication task where Alice transmits to Bob data about all scales in the theory, including scales larger than system size, which are not captured by the entanglement entropy of any subregion.

\paragraph{Overlapping $A_j$'s}
In the constrained merging protocol Alice sends data about successive regions $A_j$ to Bob. This makes sense only if the sets $A_j$ are disjoint. In more than 2+1 bulk dimensions this is not guaranteed. The large freedom in choosing shapes of boundary regions $B_j$ and $A_j$ makes it possible to construct an example, where distinct $A_j$'s overlap even though the bulk surface is convex.

\paragraph{Non-convex regions}
All our results pertain to convex surfaces.  As explained in \cite{lampros}, the differential entropy formula also computes lengths of nonconvex curves, but as a difference of two terms: integral (\ref{main}) taken over the segments where the curve is convex minus integral (\ref{main2}) taken over concave segments. Integral (\ref{main2}) showed up in Sec.~\ref{purify}, where we discussed orientation reversal and the negative ``cost'' of purifying an initial mixed state. This suggests that an information theoretic interpretation of the length of a nonconvex curve might involve a flow of information in both directions -- from Alice to Bob and from Bob to Alice. However, we have not yet succeeded in finding a well motivated quantum communication task whose cost would be precisely the length of a nonconvex curve.

\paragraph{Technical proofs}
To prove that the one-shot and von Neumann entropies coincide in the large $c$ limit, we used the eigenvalue distribution of the reduced density matrix of an interval given in \cite{calabrese-eigs}. This eigenvalue distribution applies to an interval in the vacuum or in the thermal state of a 1+1-dimensional CFT. Consequently, the proof in Sec.~\ref{sec:single-shot} is valid only for curves in pure AdS$_3$ or in the BTZ spacetime. It would be surprising, however, if analogous results did not hold in higher dimensions.

The optimality proof of Sec.~\ref{optimality} applies to the constrained merging protocol. We have not proven the optimality of the constrained swapping protocol, which is relevant to closed convex curves in AdS$_3$ and to higher-dimensional surfaces like that in Fig.~\ref{lastfig}b.

\section{Summary of results}

In this paper we have studied holographic theories of gravity, an important class of gravitational models which enjoy an equivalent description as field theories. In the last years, it has become increasingly clear that in these models the geometric structure of spacetime is intimately related to quantum information theory. In order to clarify this relation, we have given an explicit, information theoretic interpretation of one of the most basic geometric quantities in spacetime: the length of a convex curve. This interpretation involves a certain communication task in the dual field theory, whose details are encoded in the shape of the curve. Our discussion was set in the context of pure AdS$_3$, which is dual to the vacuum of a 1+1-dimensional conformal field theory. Our findings generalize to varying degrees as discussed in \secref{disc}.

The specific results are:
\begin{enumerate}
\item[Sec.~\ref{diff-entropy}:] The exhibition of a protocol for merging the state of a boundary interval from Alice to Bob at an entanglement cost equal, to leading order in the CFT central charge, to the length of a bulk curve starting and ending at the endpoints of the interval. In each step of the protocol, Alice and Bob act only in subintervals of the boundary determined by the geometry of the bulk curve. These constraints provide a precise operational implementation of the UV-IR relation.
\item[Sec.~\ref{optimality}:] A proof that, subject to appropriate locality constraints on Alice and Bob's actions, the entanglement cost is optimal: no procedure meeting the locality constraints can use less entanglement. The minimal constrained merging cost is, therefore, the length of the curve. Together, the protocol and optimality proof add a new entry to the holographic dictionary: convex bulk curves are in one-to-one correspondence with constrained boundary merging tasks whose optimal costs are the lengths of the curves themselves. From the information theory point of view, the optimality proof characterizes the rates achievable in ``streaming'' state merging protocols~\cite{blume2009streaming}.
\item[Sec.~\ref{sec:single-shot}:] A demonstration that the smooth conditional min-entropy in a 1+1 dimensional CFT with large central charge $c$ is well approximated by the conditional von Neumann entropy. As a consequence, the error terms by which the length of the curve and the entanglement cost of the communication protocol differ, vanish in the limit of large $c$.
\item[Sec.~\ref{swapprotocol}:] An analogous protocol for closed bulk curves, in which case Alice and Bob are required to swap their boundary intervals. There is at the moment, however, no matching optimality proof.
\item[Sec.~\ref{secmarkov}:] A quantitative refutation of the conjecture that the length of the bulk curve is the maximum entropy among all boundary states matching certain consistency criteria. This demonstration complements an earlier refutation~\cite{swingle-kim}. Detailed CFT calculations combined with the structure theory of quantum Markov chains reveals that the entropy of the maximally entropic consistent state is at most 2/3 of the differential entropy.
\end{enumerate}

While these results establish a clear information theoretic interpretation for the length of a curve in AdS$_3$, they leave open a number of questions. We have only worked to first order in the central charge, which we assumed was large. Quantum gravity effects typically enter as corrections to this leading order behavior, so it would be interesting to compare those corrections with more detailed calculations of the constrained merging cost to see if there is agreement. Likewise, the constrained merging interpretation proposed here depends on being able to arrange the boundary intervals in sequence, a requirement that breaks down for non-convex curves in AdS$_3$ and generically in higher dimensions. Therefore, finding an appropriate generalization of constrained state merging remains a challenge. Limitations aside, our interpretation provides a quantitative operational way of associating a bulk curve to a set of boundary degrees of freedom, helping to illuminate the meaning of holographic renormalization group flow.

\section*{Acknowledgements}
We would like to thank Lampros Lamprou, Don Marolf, Rob Myers, Steve Shenker, James Sully, and Lenny Susskind for useful discussions. BC and NL are grateful for the support from the National Science Foundation Grant No. PHYS-1066293 and the hospitality of the Aspen Center for Physics during the ``Emergent Spacetime in String Theory'' workshop. PH and BS both acknowledge support from the Simons Foundation. PH is also supported by CIFAR, FQXi and Canada's NSERC.

\appendix
\section{Saturation conditions for strong subadditivity of entropy} \label{ssa-saturation}
 Consider three disjoint systems $A$, $B$ and $C$. Strong subadditivity is the statement that
\bea\label{SSAs}
0&&\leq S(AB)+S(BC)-S(B)-S(ABC)\nn\\
&&=S(A|B)-S(A|BC)\nn\\
&&=I(A:BC)-I(A:B)=I(A:C|B).
\eea

\begin{theorem}
The following statements are equivalent: \cite{lieb-ruskai,hayden2004structure,ruskai}
\begin{itemize}
\item $\rho_{ABC}$ saturates strong subadditivity; i.e. $I(A:C|B)=0$.
\item $\log\rho_{ABC}=\log\rho_{AB}+\log\rho_{BC}-\log\rho_B$.
\item $\rho_{ABC}=\rho_{AB}^{1/2}\rho_B^{-1/2}\rho_{BC}\rho_B^{-1/2}\rho_{AB}^{1/2}$.
\item The Hilbert space of $B$ decomposes into orthogonal sectors, each admitting a decomposition into left and right degrees of freedom: $\cH_B=\oplus_i \cH_{b_L^i}\otimes \cH_{b_R^i}$ such that $\rho_{ABC}=\sum_ip_i\rho_{Ab_L^i}\otimes\rho_{b_R^iC}$.
\end{itemize}
\end{theorem}

\section{Markov operator in conformal field theory} \label{cft-calc}
In this appendix we compute the Markov operator $\sigma$ in (\ref{rho_global}) built from marginals of size $2R(x)$ in the ground state of a 1+1-dimensional conformal field theory on a line. 
The coordinate $x$ is the spatial direction of the field theory.  We have access to all marginals living on intervals $J(x)=(x-R(x),x+R(x))$.

Consider a lattice of $N$ sites at $x_i=i L/N$ for some infrared cut-off $L$. Our starting point is the vacuum density matrix of a region of size $2R(x_i)$ centered at $x_i$ in a 1+1-dimensional conformal field theory \cite{casini2011towards}:
\bea
&&\rho_{J(x_i)}=C\exp\lb -\int^{x_i+R(x_i)}_{x_i-R(x_i)} dx\:f_R(x,x_i)\:T_{00}(x)\rb,\nn\\
&&f_R(x,x_i)=2\pi \frac{R(x_i)^2-(x-x_i)^2}{2R(x_i)}.
\eea
The Markov operator $\sigma$ is defined in (\ref{rho_global}) to be
\bea
&&-\log\sigma=\sum_i\lb \log\rho_{J(x_i)}-\log\rho_{J(x_i)\cap J(x_{i+1})}\rb\nn\\
&&=\sum_i\lb\int_{x_i-R(x_i)}^{x_i+R(x_i)} dx\:f_R(x,x_i)T_{00}(x)-\int_{x_i-(R(x_i+\delta)-\delta)}^{x_i+R(x_i)}dx\:f_{R(x_i,x_{i+1})}(x,x_{i,i+1})T_{00}(x)\rb+c,\nn
\eea
where $\delta=L/N$, $R_{i,i+1}=(R(x_i)+R(x_i+\delta)-\delta)/2$ and $x_{i,i+1}=x_i+\delta/2+(R(x_i)-R(x_i+\delta))/2$.

Expanding to the first order in $\delta$ gives
\bea
-\log\sigma&&=\sum_i\int_{x_i-R(x_i)}^{x_i-(R(x_i+\delta)-\delta)} dx\:f_R(x,x_i)T_{00}(x)\nn\\
&&+\sum_i\int_{x_i-(R(x_i+\delta)-\delta)}^{x_i+R(x_i)}dx\:\lb f_R(x,x_i)-f_{R(x,x_i)}(x,x_{i,i+1})\rb T_{00}(x)+c\nn\\
&&=\sum_i\frac{\pi\delta}{2}\int_{x_i-(R(x_i+\delta)-\delta)}^{x_i+R(x_i)} dx\:\lb1-\frac{x-x_i}{R(x_i)}\rb^2(1-R'(x_i))\: T_{00}(x)+c.
\eea
Here, we have used the identities
\bea
&&\partial_Rf_R(x,x_i)=\pi\lb 1+\frac{(x-x_i)^2}{R^2}\rb,\nn\\
&&\partial_{x_i}f_R(x,x_i)=2\pi\frac{(x-x_i)}{R}.
\eea
In the limit $\delta\to 0$ the sum over intervals becomes an integral over $x_i$, and the expression for the Markov operator simplifies to
\bea
-\log\sigma&&=\frac{\pi}{2}\int_{0}^{L}dx_i\int_{x_i-R}^{x_i+R}dx\lb 1-\frac{x-x_i}{R(x_i)}\rb^2(1-R'(x_i))\: T_{00}(x)+c\nn\\
&&=\int_0^L dx\: \beta(x)T_{00}(x)+c,
\eea
where $\beta(x)=\int_{x-R}^{x+R}dx_i\lb 1-\frac{x-x_i}{R(x_i)}\rb^2(1-R'(x_i))$.
We find $\sigma$ to be proportional to a thermal density matrix with local temperature $\beta(x)$. For intervals of constant size $R(x_i)=R$  the Markov operator is the thermal state at temperature $\frac{3}{4\pi R}$:
\bea
\sigma= C e^{-\frac{4\pi R}{3} H_{CFT}},
\eea
for some constant $C$.

\section{Smooth entropy conversions} \label{smooth-ent-ineq}

The recent literature on single-shot entropies~\cite{oneshot,chain,tomamichel2010duality} usually defines smoothing with respect to the purified distance $P(\rho,\sigma)$ instead of the trace distance $T(\rho,\sigma) = \| \rho - \sigma \|_1 + |\text{tr}(\rho) - \text{tr}(\sigma)|$ that we have elected to use here (the extra term is required when considering non-normalized states). Thanks to the inequality~\cite{tomamichel2010duality}
\begin{equation} \label{P2T}
T/2 \leq P \leq \sqrt{T},
\end{equation}
approximate conversion between the two forms is straightforward. Let $\tilde H$ be the symbol for entropy smoothed with respect to $P$ rather than $T$. From \eqref{P2T} and the definition of smoothing, we immediately find
\begin{equation} \label{smooth-basics}
\tilde H_{\max}^{\sqrt{\epsilon}} \leq \Hmax^\epsilon,
\quad
\tilde H_{\max}^{\epsilon} \geq \Hmax^{\epsilon/2}
\quad \text{and} \quad
\tilde H_{\min}^{\sqrt{\epsilon}} \geq \Hmin^\epsilon.
\end{equation}
Since \cite{chain}
\begin{equation}
\tilde H_{\max}^\epsilon(A|B) \leq \tilde H_{\max}^{\epsilon/2}(AB) - \tilde H_{\min}^{\epsilon/4}(B) +\text{const},
\end{equation}
we can conclude that
\begin{equation}
\Hmax^{\epsilon}(A|B) \leq \Hmax^{\epsilon^2}(AB) - \Hmin^{\epsilon^2/4}(B) +\text{const}.
\end{equation}
Likewise, the merging bound \eqref{ub-single-shot} was originally stated in terms of $- \tilde H_{\min}^{\epsilon^2/13}(A|R)$~\cite{oneshot}. The virtue of the purified distance, however, is that it obeys a convenient relationship between min- and max-entropies: for a pure state $\ket{\ph}_{ABR}$, semidefinite programming duality can be used to show that  $-\tilde H_{\min}^{\epsilon^2/13}(A|R) = \tilde H_{\max}^{\epsilon^2/13}(A|B)$~\cite{tomamichel2010duality}. The latter is bounded above by $\Hmax^{\epsilon^4/169}(A|B)$ by \eqref{smooth-basics}.

In fact, the purified distance can be computed in our setting. Given two positive operators $\rho$ and $\sigma$ with at least one of them normalized, the purified distance $P(\rho,\sigma)$ is defined in terms of the fidelity $F(\rho,\sigma)$ as
\beq
P = \sqrt{1-F^2}
\eeq
with
\beq
F = \| \sqrt{\sqrt{\rho}\sigma \sqrt{\rho}}\|_1.
\eeq
Supposing that $\rho$ is the (normalized) density operator of a given interval and $\sigma$ is its truncation, it follows that $[\rho,\sigma]=0$ which vastly simplifies the fidelity. In this case
\beq
F = \| \sqrt{\rho \sigma} \|_1 = \| \sigma \|_1 = 1-\epsilon,
\eeq
where the first equality is the definition, the second follows from the fact that $\rho \sigma = \sigma^2$, and the final equality is part of the definition of $\sigma$. Plugging this form into $P$ yields
\beq
P(\rho,\sigma) = \sqrt{1-F^2} = \sqrt{2 \epsilon - \epsilon^2} = \sqrt{2 \epsilon}(1+ O(\epsilon)),
\eeq
and since $T(\rho,\sigma) = \| \rho - \sigma \|_1 + |\text{tr}(\rho) - \text{tr}(\sigma)| = 2 \epsilon$ the upper bound in \eqref{P2T} is almost saturated.

\section{Norm of the Markov operator} \label{golden}
\begin{theorem}
Consider a global state $\rho$ on a line and its marginals on a set of intervals $I_j$ that we denote by $\rho_{I_j}$. Assume that for all $j$, $I_j$ is to the right of $I_{j-1}$, that is
\bea
\forall j:\:\emptyset\neq I_j\cap I_{j-1}=I_j\cap (\cup_{i=1}^{j-1} I_i).
\eea
Then,
\bea
\tr \exp\lb\sum_j \log\rho_{I_j}-\log\rho_{I_j\cap I_{j-1}}\rb\leq 1.
\eea
If all $\rho_{I_j}$ and $\rho_{I_j\cap I_{j-1}}$ commute, then the inequality is saturated.
\end{theorem}
Consider the first three intervals $I_1$, $I_2$ and $I_3$. Since $\log\rho$ is a Hermitian and bounded operator for all intervals we have \cite{lieb1973convex}
\bea
\tr \sigma_{12}&\equiv&\exp\lb \log\rho_{I_1}+\log\rho_{I_2}-\log\rho_{I_2\cap I_1}\rb\nn\\
&\leq& \tr\lb \int_0^\infty \rho_{I_1}(\rho_{I_2\cap I_1}+x \mathbb{I})^{-1}\rho_{I_2}(\rho_{I_2\cap I_1}+x \mathbb{I})^{-1}dx\rb\nn\\
&=&\tr\lb \int_0^\infty \rho_{I_2\cap I_1}(\rho_{I_2\cap I_1}+x \mathbb{I})^{-1}\rho_{I_2\cap I_1}(\rho_{I_2\cap I_1}+x \mathbb{I})^{-1}dx\rb\nn\\
&=&\tr\rho_{I_2\cap I_1}=1.
\eea
The inequality is saturated if $\rho_{I_1}$, $\rho_{I_2}$ and $\rho_{I_1\cap I_2}$ commute. Since $\sigma_{12}$ is also bounded and Hermitian, we can rewrite the same inequality for $\log\sigma_{12}$, $\log\rho_{I_3}$ and $\log\rho_{I_2\cap I_3}$ to find
\bea
\tr\sigma_{123}&\equiv&\exp\lb \log\sigma_{12}+\log\rho_{I_3}-\log\rho_{I_3\cap I_2}\rb\nn\\
&\leq& \tr\lb \int_0^\infty \sigma_{12}(\rho_{I_3\cap I_2}+x \mathbb{I})^{-1}\rho_{I_3}(\rho_{I_3\cap I_2}+x \mathbb{I})^{-1}dx\rb\nn\\
&=&\tr\lb \int_0^\infty \sigma_{12}(\rho_{I_3\cap I_2}+x \mathbb{I})^{-1}\rho_{I_3\cap I_2}(\rho_{I_3\cap I_2}+x \mathbb{I})^{-1}dx\rb\nn\\
&=&\tr\sigma_{12}\leq 1,
\eea
with the equality condition that $\rho_{I_3}$, $\rho_{I_2}$ and $\sigma_{12}$ commute. Repeating this inequality $N$ times we find that the trace of the Markov operator is less than $1$:
\bea\label{trsigma}
\tr\sigma&\equiv &\tr\exp\lb\log\rho_{I_1}+\sum_{j=2}^{N+2}\log \rho_{I_j}-\log\rho_{I_j\cap I_{j-1}}\rb\leq 1.
\eea
The inequality in (\ref{trsigma}) is an equality if all $\rho_{I_j}$ and $\rho_{I_{j,j-1}}$ commute. On a circle there is an extra term in the exponent of $\sigma$ with a negative sign. Unfortunately, we do not know how to generalize our argument to apply to this case.

\bibliographystyle{JHEP}

\bibliography{diffprotocol}

\providecommand{\href}[2]{#2}\begingroup\raggedright\begin{thebibliography}{10}

\bibitem{adscft}
J.~{Maldacena}, {\it {The Large-N Limit of Superconformal Field Theories and
  Supergravity}},  {\em International Journal of Theoretical Physics} {\bf 38}
  (1999) 1113--1133, [\href{http://arxiv.org/abs/hep-th/9711200}{{\tt
  hep-th/9711200}}].

\bibitem{witten98}
E.~Witten, {\it {Anti-de Sitter space and holography}},  {\em
  Adv.Theor.Math.Phys.} {\bf 2} (1998) 253--291,
  [\href{http://arxiv.org/abs/hep-th/9802150}{{\tt hep-th/9802150}}].

\bibitem{maldacena2003tasi}
J.~M. Maldacena, {\it Tasi 2003 lectures on ads/cft},  {\em arXiv preprint
  hep-th/0309246} {\bf 4} (2003).

\bibitem{hartnoll2009lectures}
S.~A. Hartnoll, {\it Lectures on holographic methods for condensed matter
  physics},  {\em Classical and Quantum Gravity} {\bf 26} (2009), no.~22
  224002.

\bibitem{2005PhRvL..94k1601K}
P.~K. {Kovtun}, D.~T. {Son}, and A.~O. {Starinets}, {\it {Viscosity in Strongly
  Interacting Quantum Field Theories from Black Hole Physics}},  {\em Physical
  Review Letters} {\bf 94} (Mar., 2005) 111601,
  [\href{http://arxiv.org/abs/hep-th/0405231}{{\tt hep-th/0405231}}].

\bibitem{2007PhRvD..75h5020H}
C.~P. {Herzog}, P.~{Kovtun}, S.~{Sachdev}, and D.~T. {Son}, {\it {Quantum
  critical transport, duality, and M theory}},  {\em Physics Review D} {\bf 75}
  (Apr., 2007) 085020, [\href{http://arxiv.org/abs/hep-th/0701036}{{\tt
  hep-th/0701036}}].

\bibitem{myreview}
V.~Balasubramanian and B.~Czech, {\it {Quantitative approaches to information
  recovery from black holes}},  {\em Class.Quant.Grav.} {\bf 28} (2011) 163001,
  [\href{http://arxiv.org/abs/1102.3566}{{\tt arXiv:1102.3566}}].

\bibitem{tedsessay}
T.~Jacobson, {\it {Boundary unitarity and the black hole information paradox}},
   {\em Int.J.Mod.Phys.} {\bf D22} (2013) 1342002,
  [\href{http://arxiv.org/abs/1212.6944}{{\tt arXiv:1212.6944}}].

\bibitem{uvir}
L.~Susskind and E.~Witten, {\it {The Holographic bound in anti-de Sitter
  space}},  \href{http://arxiv.org/abs/hep-th/9805114}{{\tt hep-th/9805114}}.

\bibitem{Akhmedov:1998vf}
E.~T. Akhmedov, {\it {A Remark on the AdS / CFT correspondence and the
  renormalization group flow}},  {\em Phys.Lett.} {\bf B442} (1998) 152--158,
  [\href{http://arxiv.org/abs/hep-th/9806217}{{\tt hep-th/9806217}}].

\bibitem{Balasubramanian:1999jd}
V.~Balasubramanian and P.~Kraus, {\it {Space-time and the holographic
  renormalization group}},  {\em Phys.Rev.Lett.} {\bf 83} (1999) 3605--3608,
  [\href{http://arxiv.org/abs/hep-th/9903190}{{\tt hep-th/9903190}}].

\bibitem{deBoer:1999xf}
J.~de~Boer, E.~P. Verlinde, and H.~L. Verlinde, {\it {On the holographic
  renormalization group}},  {\em JHEP} {\bf 0008} (2000) 003,
  [\href{http://arxiv.org/abs/hep-th/9912012}{{\tt hep-th/9912012}}].

\bibitem{Balasubramanian:2012hb}
V.~Balasubramanian, M.~Guica, and A.~Lawrence, {\it {Holographic
  Interpretations of the Renormalization Group}},  {\em JHEP} {\bf 1301} (2013)
  115, [\href{http://arxiv.org/abs/1211.1729}{{\tt arXiv:1211.1729}}].

\bibitem{Jackson:2013eqa}
S.~Jackson, R.~Pourhasan, and H.~Verlinde, {\it {Geometric RG Flow}},
  \href{http://arxiv.org/abs/1312.6914}{{\tt arXiv:1312.6914}}.

\bibitem{rt1}
S.~{Ryu} and T.~{Takayanagi}, {\it {Holographic Derivation of Entanglement
  Entropy from the anti de Sitter Space/Conformal Field Theory
  Correspondence}},  {\em Physical Review Letters} {\bf 96} (May, 2006) 181602,
  [\href{http://arxiv.org/abs/hep-th/0603001}{{\tt hep-th/0603001}}].

\bibitem{rt2}
S.~{Ryu} and T.~{Takayanagi}, {\it {Aspects of holographic entanglement
  entropy}},  {\em Journal of High Energy Physics} {\bf 8} (Aug., 2006) 45,
  [\href{http://arxiv.org/abs/hep-th/0605073}{{\tt hep-th/0605073}}].

\bibitem{entcorr}
M.~M. Wolf, F.~Verstraete, M.~B. Hastings, and J.~I. Cirac, {\it Area laws in
  quantum systems: Mutual information and correlations},  {\em Phys. Rev.
  Lett.} {\bf 100} (Feb, 2008) 070502.

\bibitem{Horowitz:1998xk}
G.~T. Horowitz and D.~Marolf, {\it {A New approach to string cosmology}},  {\em
  JHEP} {\bf 9807} (1998) 014, [\href{http://arxiv.org/abs/hep-th/9805207}{{\tt
  hep-th/9805207}}].

\bibitem{Balasubramanian:1998de}
V.~Balasubramanian, P.~Kraus, A.~E. Lawrence, and S.~P. Trivedi, {\it
  {Holographic probes of anti-de Sitter space-times}},  {\em Phys.Rev.} {\bf
  D59} (1999) 104021, [\href{http://arxiv.org/abs/hep-th/9808017}{{\tt
  hep-th/9808017}}].

\bibitem{Maldacena:2001kr}
J.~M. Maldacena, {\it {Eternal black holes in anti-de Sitter}},  {\em JHEP}
  {\bf 0304} (2003) 021, [\href{http://arxiv.org/abs/hep-th/0106112}{{\tt
  hep-th/0106112}}].

\bibitem{markessay}
M.~{van Raamsdonk}, {\it {Building up spacetime with quantum entanglement}},
  {\em General Relativity and Gravitation} {\bf 42} (Oct., 2010) 2323--2329,
  [\href{http://arxiv.org/abs/1005.3035}{{\tt arXiv:1005.3035}}].

\bibitem{rqg}
B.~Czech, J.~L. Karczmarek, F.~Nogueira, and M.~Van~Raamsdonk, {\it {Rindler
  Quantum Gravity}},  {\em Class.Quant.Grav.} {\bf 29} (2012) 235025,
  [\href{http://arxiv.org/abs/1206.1323}{{\tt arXiv:1206.1323}}].

\bibitem{balasubramanian2014multiboundary}
V.~Balasubramanian, P.~Hayden, A.~Maloney, D.~Marolf, and S.~F. Ross, {\it
  Multiboundary wormholes and holographic entanglement},  {\em Classical and
  Quantum Gravity} {\bf 31} (2014), no.~18 185015.

\bibitem{ssaproof}
M.~Headrick and T.~Takayanagi, {\it {A Holographic proof of the strong
  subadditivity of entanglement entropy}},  {\em Phys.Rev.} {\bf D76} (2007)
  106013, [\href{http://arxiv.org/abs/0704.3719}{{\tt arXiv:0704.3719}}].

\bibitem{lampros}
B.~Czech and L.~Lamprou, {\it {Nuts and Bolts for Creating Space}},
  \href{http://arxiv.org/abs/1409.4473}{{\tt arXiv:1409.4473}}.

\bibitem{ssalorentz}
H.~Casini and M.~Huerta, {\it {A c-theorem for the entanglement entropy}},
  {\em J.Phys.} {\bf A40} (2007) 7031--7036,
  [\href{http://arxiv.org/abs/cond-mat/0610375}{{\tt cond-mat/0610375}}].

\bibitem{monogamy}
P.~Hayden, M.~Headrick, and A.~Maloney, {\it {Holographic Mutual Information is
  Monogamous}},  {\em Phys.Rev.} {\bf D87} (2013), no.~4 046003,
  [\href{http://arxiv.org/abs/1107.2940}{{\tt arXiv:1107.2940}}].

\bibitem{briansessay}
B.~Swingle, {\it {Entanglement Renormalization and Holography}},  {\em
  Phys.Rev.} {\bf D86} (2012) 065007,
  [\href{http://arxiv.org/abs/0905.1317}{{\tt arXiv:0905.1317}}].

\bibitem{marks1st}
M.~Van~Raamsdonk, {\it {Comments on quantum gravity and entanglement}},
  \href{http://arxiv.org/abs/0907.2939}{{\tt arXiv:0907.2939}}.

\bibitem{brians2nd}
B.~Swingle, {\it {Constructing holographic spacetimes using entanglement
  renormalization}},  \href{http://arxiv.org/abs/1209.3304}{{\tt
  arXiv:1209.3304}}.

\bibitem{bianchimyers}
E.~Bianchi and R.~C. Myers, {\it {On the Architecture of Spacetime Geometry}},
  \href{http://arxiv.org/abs/1212.5183}{{\tt arXiv:1212.5183}}.

\bibitem{tomjuan}
T.~Hartman and J.~Maldacena, {\it {Time Evolution of Entanglement Entropy from
  Black Hole Interiors}},  {\em JHEP} {\bf 1305} (2013) 014,
  [\href{http://arxiv.org/abs/1303.1080}{{\tt arXiv:1303.1080}}].

\bibitem{myerssmolkin}
R.~C. Myers, R.~Pourhasan, and M.~Smolkin, {\it {On Spacetime Entanglement}},
  {\em JHEP} {\bf 1306} (2013) 013, [\href{http://arxiv.org/abs/1304.2030}{{\tt
  arXiv:1304.2030}}].

\bibitem{holeentropy}
V.~Balasubramanian, B.~Czech, B.~D. Chowdhury, and J.~de~Boer, {\it {The
  entropy of a hole in spacetime}},  {\em JHEP} {\bf 1310} (2013) 220,
  [\href{http://arxiv.org/abs/1305.0856}{{\tt arXiv:1305.0856}}].

\bibitem{erepr}
J.~Maldacena and L.~Susskind, {\it {Cool horizons for entangled black holes}},
  {\em Fortsch.Phys.} {\bf 61} (2013) 781--811,
  [\href{http://arxiv.org/abs/1306.0533}{{\tt arXiv:1306.0533}}].

\bibitem{xiaoliang}
X.-L. Qi, {\it {Exact holographic mapping and emergent space-time geometry}},
  \href{http://arxiv.org/abs/1309.6282}{{\tt arXiv:1309.6282}}.

\bibitem{complexity}
D.~Stanford and L.~Susskind, {\it {Complexity and Shock Wave Geometries}},
  \href{http://arxiv.org/abs/1406.2678}{{\tt arXiv:1406.2678}}.

\bibitem{holeography}
V.~{Balasubramanian}, B.~D. {Chowdhury}, B.~{Czech}, J.~{de Boer}, and M.~P.
  {Heller}, {\it {A hole-ographic spacetime}},  {\em Phys.~Rev.~D} {\bf 89}
  (Apr., 2014) 086004, [\href{http://arxiv.org/abs/1310.4204}{{\tt
  arXiv:1310.4204}}].

\bibitem{naturepaper}
M.~{Horodecki}, J.~{Oppenheim}, and A.~{Winter}, {\it {Partial quantum
  information}},  {\em Nature} {\bf 436} (Aug., 2005) 673--676,
  [\href{http://arxiv.org/abs/quant-ph/0505062}{{\tt quant-ph/0505062}}].

\bibitem{horodecki2007quantum}
M.~Horodecki, J.~Oppenheim, and A.~Winter, {\it Quantum state merging and
  negative information},  {\em Communications in Mathematical Physics} {\bf
  269} (2007), no.~1 107--136.

\bibitem{schumacher}
B.~Schumacher, {\it Quantum coding},  {\em Phys. Rev. A} {\bf 51} (Apr, 1995)
  2738--2747.

\bibitem{teleport}
C.~H. Bennett, G.~Brassard, C.~Cr\'epeau, R.~Jozsa, A.~Peres, and W.~K.
  Wootters, {\it Teleporting an unknown quantum state via dual classical and
  einstein-podolsky-rosen channels},  {\em Phys. Rev. Lett.} {\bf 70} (Mar,
  1993) 1895--1899.

\bibitem{brownhen}
J.~D. Brown and M.~Henneaux, {\it Central charges in the canonical realization
  of asymptotic symmetries: an example from three-dimensional gravity},  {\em
  Communications in Mathematical Physics} {\bf 104} (1986), no.~2 207--226.

\bibitem{roblast}
R.~C. {Myers}, J.~{Rao}, and S.~{Sugishita}, {\it {Holographic holes in higher
  dimensions}},  {\em Journal of High Energy Physics} {\bf 6} (June, 2014) 44,
  [\href{http://arxiv.org/abs/1403.3416}{{\tt arXiv:1403.3416}}].

\bibitem{xi}
B.~{Czech}, X.~{Dong}, and J.~{Sully}, {\it {Holographic Reconstruction of
  General Bulk Surfaces}},  {\em ArXiv e-prints} (June, 2014)
  [\href{http://arxiv.org/abs/1406.4889}{{\tt arXiv:1406.4889}}].

\bibitem{entwinement}
V.~{Balasubramanian}, B.~D. {Chowdhury}, B.~{Czech}, and J.~{de Boer}, {\it
  {Entwinement and the emergence of spacetime}},  {\em ArXiv e-prints} (June,
  2014) [\href{http://arxiv.org/abs/1406.5859}{{\tt arXiv:1406.5859}}].

\bibitem{Wienthesis}
J.~Wien, {\it {A Holographic Approach to Spacetime Entanglement}},
  \href{http://arxiv.org/abs/1408.6005}{{\tt arXiv:1408.6005}}.

\bibitem{robproof}
M.~Headrick, R.~C. Myers, and J.~Wien, {\it {Holographic Holes and Differential
  Entropy}},  \href{http://arxiv.org/abs/1408.4770}{{\tt arXiv:1408.4770}}.

\bibitem{bennett1996mixed}
C.~H. Bennett, D.~P. DiVincenzo, J.~A. Smolin, and W.~K. Wootters, {\it
  Mixed-state entanglement and quantum error correction},  {\em Physical Review
  A} {\bf 54} (1996), no.~5 3824.

\bibitem{devetak2005distillation}
I.~Devetak and A.~Winter, {\it Distillation of secret key and entanglement from
  quantum states},  {\em Proceedings of the Royal Society A: Mathematical,
  Physical and Engineering Science} {\bf 461} (2005), no.~2053 207--235.

\bibitem{uhlmann1976transition}
A.~Uhlmann, {\it The “transition probability” in the state space of
  a∗-algebra},  {\em Reports on Mathematical Physics} {\bf 9} (1976), no.~2
  273--279.

\bibitem{fuchs}
C.~A. Fuchs and J.~van~de Graaf, {\it Cryptographic distinguishability measures
  for quantum-mechanical states},  {\em {IEEE} Transactions on Information
  Theory} {\bf 45} (1999) 1216.

\bibitem{alicki2004continuity}
R.~Alicki and M.~Fannes, {\it Continuity of quantum conditional information},
  {\em Journal of Physics {A}: mathematical and general} {\bf 37} (2004), no.~5
  L55--L57.

\bibitem{oneshot}
F.~Dupuis, M.~Berta, J.~Wullschleger, and R.~Renner, {\it One-shot decoupling},
   {\em Communications in Mathematical Physics} {\bf 328} (2014), no.~1
  251--284.

\bibitem{chain}
A.~Vitanov, F.~Dupuis, M.~Tomamichel, and R.~Renner, {\it Chain rules for
  smooth min- and max-entropies},  {\em {IEEE} Transactions on Information
  Theory} {\bf 59} (2013), no.~5 2603--2612.

\bibitem{calabrese-eigs}
P.~{Calabrese} and A.~{Lefevre}, {\it {Entanglement spectrum in one-dimensional
  systems}},  {\em Phys.~Rev.~A} {\bf 78} (Sept., 2008) 032329,
  [\href{http://arxiv.org/abs/0806.3059}{{\tt arXiv:0806.3059}}].

\bibitem{veronikajune}
V.~E. Hubeny, {\it {Covariant Residual Entropy}},
  \href{http://arxiv.org/abs/1406.4611}{{\tt arXiv:1406.4611}}.

\bibitem{swingle-kim}
B.~{Swingle} and I.~H. {Kim}, {\it {Reconstructing quantum states from local
  data}},  {\em ArXiv e-prints} (July, 2014)
  [\href{http://arxiv.org/abs/1407.2658}{{\tt arXiv:1407.2658}}].

\bibitem{poulin2011markov}
D.~Poulin and M.~B. Hastings, {\it Markov entropy decomposition: a variational
  dual for quantum belief propagation},  {\em Physical review letters} {\bf
  106} (2011), no.~8 080403.

\bibitem{liu2006consistency}
Y.-K. Liu, {\it Consistency of local density matrices is qma-complete},  in
  {\em Approximation, Randomization, and Combinatorial Optimization. Algorithms
  and Techniques}, pp.~438--449.
\newblock Springer, 2006.

\bibitem{hayden2004structure}
P.~Hayden, R.~Jozsa, D.~Petz, and A.~Winter, {\it Structure of states which
  satisfy strong subadditivity of quantum entropy with equality},  {\em
  Communications in mathematical physics} {\bf 246} (2004), no.~2 359--374.

\bibitem{fawzi2014quantum}
O.~Fawzi and R.~Renner, {\it Quantum conditional mutual information and
  approximate quantum {M}arkov chains},  {\em ar{X}iv:1410.0664} (2014).

\bibitem{ruskai}
M.~B. Ruskai, {\it Inequalities for quantum entropy: a review with conditions
  for equality},  {\em Journal of Mathematical Physics} {\bf 43} (2002)
  4358--4375.

\bibitem{casini2011towards}
H.~Casini, M.~Huerta, and R.~C. Myers, {\it Towards a derivation of holographic
  entanglement entropy},  {\em Journal of High Energy Physics} {\bf 2011}
  (2011), no.~5 1--41.

\bibitem{schwinger1963commutation}
J.~Schwinger, {\it Commutation relations and conservation laws},  {\em Physical
  Review} {\bf 130} (1963), no.~1 406.

\bibitem{Nogueira:2013if}
F.~Nogueira, {\it {Extremal Surfaces in Asymptotically AdS Charged Boson Stars
  Backgrounds}},  {\em Phys.Rev.} {\bf D87} (2013), no.~10 106006,
  [\href{http://arxiv.org/abs/1301.4316}{{\tt arXiv:1301.4316}}].

\bibitem{Gentle:2013fma}
S.~A. Gentle and M.~Rangamani, {\it {Holographic entanglement and causal
  information in coherent states}},  {\em JHEP} {\bf 1401} (2014) 120,
  [\href{http://arxiv.org/abs/1311.0015}{{\tt arXiv:1311.0015}}].

\bibitem{renyis}
M.~Headrick, {\it {Entanglement Renyi entropies in holographic theories}},
  {\em Phys.Rev.} {\bf D82} (2010) 126010,
  [\href{http://arxiv.org/abs/1006.0047}{{\tt arXiv:1006.0047}}].

\bibitem{Hubeny:2012ry}
V.~E. Hubeny, {\it {Extremal surfaces as bulk probes in AdS/CFT}},  {\em JHEP}
  {\bf 1207} (2012) 093, [\href{http://arxiv.org/abs/1203.1044}{{\tt
  arXiv:1203.1044}}].

\bibitem{plateaux}
V.~E. Hubeny, H.~Maxfield, M.~Rangamani, and E.~Tonni, {\it {Holographic
  entanglement plateaux}},  {\em JHEP} {\bf 1308} (2013) 092,
  [\href{http://arxiv.org/abs/1306.4004}{{\tt arXiv:1306.4004}}].

\bibitem{blume2009streaming}
R.~Blume-Kohout, S.~Croke, and D.~Gottesman, {\it Streaming universal
  distortion-free entanglement concentration},  {\em ar{X}iv:0910.5952} (2009).

\bibitem{lieb-ruskai}
E.~H. Lieb and M.~B. Ruskai, {\it Proof of the strong subadditivity of
  quantum‐mechanical entropy},  {\em Journal of Mathematical Physics} {\bf
  14} (1973), no.~12 1938--1941.

\bibitem{tomamichel2010duality}
M.~Tomamichel, R.~Colbeck, and R.~Renner, {\it Duality between smooth min-and
  max-entropies},  {\em Information Theory, IEEE Transactions on} {\bf 56}
  (2010), no.~9 4674--4681.

\bibitem{lieb1973convex}
E.~H. Lieb, {\it Convex trace functions and the wigner-yanase-dyson
  conjecture},  {\em Advances in Mathematics} {\bf 11} (1973), no.~3 267--288.

\end{thebibliography}\endgroup

\end{document}